# Can large language models generate geospatial code?

Shuyang Hou[a], Zhangxiao Shen[a], Jianyuan Liang[a], Anqi Zhao[a], Zhipeng Gui[b], Rui Li[a], Huayi Wu[a]*

*a. State Key Laboratory of Information Engineering in Surveying, Mapping and Remote Sensing, Wuhan University, Wuhan, China
b. School ofRemote Sensing and Information Engineering, Wuhan University
Wuhan, China
*Corresponding author: Huayi Wu, email: wuhuayi@whu.edu.cn

**Abstract:** With the growing demand for spatiotemporal data processing and geospatial modeling, the ability to automatically generate geospatial code has become a critical way to enhance productivity. Large language models (LLMs) have shown considerable potential in code generation tasks. However, the lack of domain-specific knowledge and code corpora often leads to challenges such as refusal to generate or "coding hallucinations". The capacity of LLMs to generate effective geospatial code remains uncertain, thereby impeding advancements in research and applications. To address this gap, this paper introduces an evaluation framework for assessing the geospatial code generation capabilities of LLMs. The proposed GeoCode-Eval (GCE) framework evaluates three dimensions: "Cognition and Memory", "Comprehension and Interpretation", and "Innovation and Creation", distributed across eight distinct capability levels. We developed a benchmark dataset, GeoCode-Bench, comprising 5,000 multiple-choice questions, 1,500 fill-in-the-blank questions, 1,500 true/false questions, and 1,000 subjective tasks in code summarization, code generation, code completion, and code correction. Using GeoCode-Bench, we conducted a comprehensive evaluation of the geospatial code generation capabilities of 3 commercial closed-source LLMs, 4 open-source general-purpose LLMs, and 14 specialized code generation models. Additionally, we conducted contrastive experiments involving various prompting strategies, such as few-shot and zero-shot learning, Chain of Thought (CoT) reasoning, and multi-round majority voting, to analyze their impact on the geospatial code generation performance of LLMs. To further evaluate the effectiveness of fine-tuning techniques for geospatial code generation, we fine-tuned the Code LLaMA-7B model with JavaScript scripts and related task instructions from the Google Earth Engine platform, creating the GEECode-GPT model. This fine-tuned model was subsequently evaluated on subjective tasks. Our experimental results indicate that constructing small-scale pre-training and instruction datasets can substantially improve the model's code generation performance, offering valuable insights for optimizing LLMs in specific domains.

**Keywords:** code generation, geospatial code, large language models, fine-tuning, Google Earth Engine

# 1  Introduction

General-purpose large language models (LLMs) have demonstrated exceptional capabilities in contextual learning, reasoning, and executing instructions thanks to the training process on diverse corpora including articles, web pages, scientific databases, books, news, and code[1]. Due to the relatively small proportion of code in the training data[2], these models often face challenges such as refusal to generate code or "coding hallucinations" in code generation tasks[3]. By fine-tuning models with general-purpose code across various programming languages and coding styles, as well as related task instructions, specialized models for code generation can be constructed[4]. However, their capabilities rely on clear input-output instructions and primarily limited in basic modalities such as characters, numerical values, bitmaps, and graph structures[5].

With the intelligent transformation of various fields, analysis platforms and function toolkits developed with general-purpose programming languages have emerged across specialized fields[6]. For instance, in life sciences, Bioconductor provides R packages for genomic data analysis[7]; in finance, QuantLib offers C++ toolkits for pricing financial derivatives[8]; and in chemistry, RDKit uses Python for cheminformatics and molecular modeling[9]. These platforms or toolkits are developed with general-purpose programming languages, integrating internal datasets and custom parameters[10], having specific syntax rules and data processing workflows to handle multimodal data (e.g., multispectral remote sensing imagery or molecular structure data in life sciences). Their data structures can be complex, differing significantly from general-purpose code in both syntax and objects of processing[11].

The geospatial domain is distinguished as spatiotemporal datasets with vast volume and the integration of multiple modalities, including precise longitude and latitude, surface elevation, geographic topology, and multispectral remote sensing imagery[12]. The effective management and processing of such complex datasets necessitate the use of geospatial tools and platforms, including geospatial cloud computing frameworks like Google Earth Engine and PIE Engine, as well as computational libraries such as Python's GeoPandas, ArcPy, and R's Raster and Terra. Each of these platforms and tools has specialized operator structures and function syntaxes, which contribute to significant learning barriers and hinder the efficient development of geospatial processing code[13]. Consequently, the demand for automated geospatial code generation has intensified, aiming to enhance modeling efficiency and facilitate greater productivity in geospatial research and practical applications[14].

This raises an important question: **Can LLMs trained on general-purpose corpora effectively acquire the domain-specific knowledge for geospatial code generation and produce high-quality code?** To address this question, this paper proposes a **GeoCode-Eval (GCE)** framework, based on the cognitive taxonomy theory of American educational psychologist Benjamin Samuel Bloom[15]. The GCE framework evaluates geospatial code generation across three dimensions—"Cognition and Memory", "Comprehension and Interpretation", and "Innovation and Creation"—as shown in Figure 1. The **Cognition and Memory** dimension assesses: (1) whether the model has basic knowledge of geospatial analysis platforms or toolkits; (2) whether it understands functions related to geospatial analysis; and (3) whether it is familiar with the built-in datasets of these platforms. The **Comprehension and Interpretation** dimension further evaluates: (4) whether the model can accurately identify key entities in geospatial code; and (5) whether it can generate correct code summaries. The **Innovation and Creation** dimension assesses: (6) whether the model can generate complete geospatial code according to instructions; (7) whether it can effectively complete code; and (8) whether it can correct geospatial code based on given instructions.

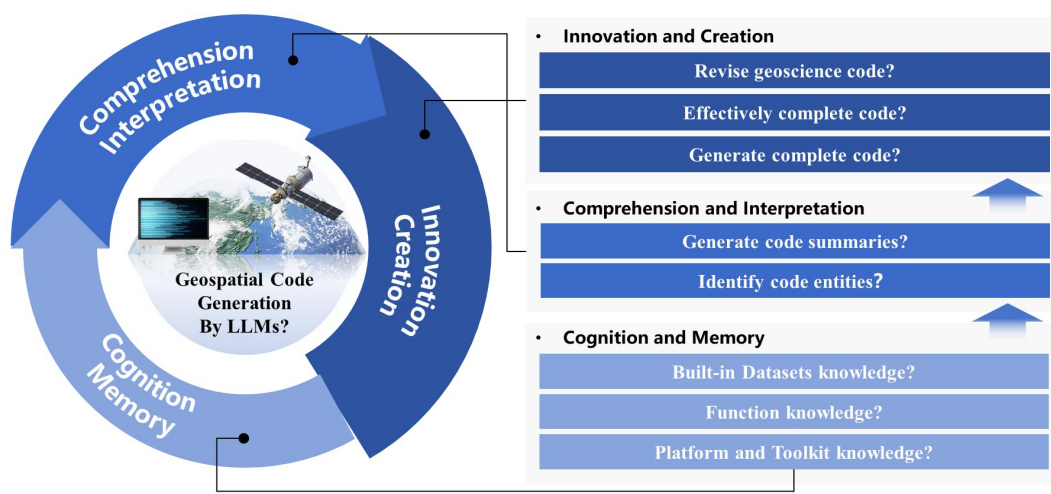

Figure.1 GeoCode-Eval (GCE) Evaluation Framework Diagram

Around the GCE evaluation framework, we collected a large amount of real-world geospatial code, related Wikipedia pages and introduction pages, information on built-in functions, syntax, and datasets from geospatial analysis platforms and toolkits from open-source platforms. Using expert experience, the Self-Instruct framework, and exhaustive traversal methods of open-source platforms, we constructed the **GeoCode-Bench** benchmark dataset. The dataset covers seven types of questions: three objective categories—multiple-choice, fill-in-the-blank, and true/false questions—as well as four subjective categories, including code summarization, code generation, code completion, and code correction. All questions were revised and reviewed by experts. The evaluated models include leading commercial LLMs, general-purpose LLMs, and code generation models, with different parameter scales of some models chosen as comparisons, aiming to comprehensively evaluate the capacity boundaries of these models. The evaluation strategy employed a combination of expert assessments and LLM prompt engineering. Multiple answering methods were employed, including multi-round voting, few-shot prompting (1-shot), Chain of Thought (CoT) prompting, and direct zero-shot (0-shot) prompting, to evaluate the impact of different prompting methods on geospatial code generation performance, followed by data analysis and visualization. Finally, this paper proposes enhancing the performance of LLMs in geospatial code generation through instruction fine-tuning. Specifically, we built pre-training and instruction datasets from open-source platforms and fine-tuned the Code Llama-7B model to obtain GEECode-GPT. Compared to the non-fine-tuned model, this approach significantly improved its ability to generate geospatial code in the Google Earth Engine platform's JavaScript environment.

The contributions of this paper can be summarized as follows:

• This paper introduces a novel evaluation task for LLMs, specifically assessing their capabilities in geospatial code generation. The proposed GeoCode-Eval (GCE) framework is adaptable to other evaluation tasks for LLMs across different domains.

• We constructed the GeoCode-Bench benchmark dataset, providing a standardized tool for evaluating the geospatial code generation capabilities of various LLMs.

• This paper provides the first comprehensive evaluation of LLM performance in geospatial code generation from multiple perspectives, including model category, parameter scales, and prompting strategies, offering detailed data to support future downstream tasks.

• A novel technical approach for enhancing geospatial code generation capabilities in LLMs through fine-tuning is proposed. The comparative analysis demonstrates significant improvements in geospatial code generation performance, offering insights for future optimization of LLMs in specialized domains.

The remainder of this paper is organized as follows: Section 2 reviews related works on LLMs regarding performance variability, code generation, performance evaluation, and improvement methods. Section 3 details the construction of the GeoCode-Bench benchmark dataset. Section 4 describes the evaluation experiment design, the evaluated models, the evaluation strategies, and evaluation results. Section 5 presents case studies on constructing pre-training and instruction datasets through JavaScript code on the GEE platform and improving LLM geospatial code generation capabilities through fine-tuning. Finally, Section 6 concludes the paper and discusses future research directions.

## 2 Related Works

### 2.1 Differences Among Large Language Models

The performance of LLMs in contextual learning, reasoning, and instruction execution is primarily influenced by various factors including corpus size and diversity[16], model parameters[17], training strategies[18], prompting techniques[19], and inference hardware[20]. A broader corpus generally results in better model generalization, but in specialized domains, a customized corpus with precise domain knowledge can be more effective. Larger scale models tend to perform better, but with lower inference efficiency. Different training optimizations and prompting strategies—such as zero-shot learning[21], prompt engineering[22], and Chain of Thought reasoning[23]—also significantly affect output quality. Open-source models are more cost-effective for inference but may have inferior performance compared to proprietary models. Sufficient computational resources are also crucial for model inference. Therefore, evaluating LLM capabilities in specific domains can help improve model training, inference configurations, and prompting strategies[1].

### 2.2 Code Generation with Large Language Models

The vast corpus and large parameter scale are both an advantage and a burden for

LLMs[24]: retraining requires substantial computational resources, but knowledge gaps are likely to occur without timely retraining to update knowledge[25]. The general corpus is overly broad, but when handling specialized tasks, models are prone to "knowledge hallucinations[26]", generating incorrect answers. To address this, promoting domain specialization for LLMs (known as LLM+X paradigm) has become a research focus[27], showing effectiveness in the fields of social sciences[28-30] and natural sciences[31-33]. Code generation tasks which convert natural language instructions into executable code, is an important direction for LLM domain specialization[5]. Current general-purpose code generation models, such as Code LLaMA[34], DeepSeek Coder[35] and Code Qwen[36], have been widely used, but models specialized for geospatial code generation have not yet emerged[37].

## 2.3 Evaluation of Large Language Model Capabilities

Both academia and industry are exploring how LLMs can disrupt existing research paradigms and are investigating their capacity and application potential. The evaluation of LLM capabilities across various dimensions has become a research hotspot[1], encompassing domains such as simulation of human cognitive behavior[38-44], mathematics and data comprehension[45-49], social sciences[50], medical and healthcare[51-56], legal and ethical[57], education[58-60], software security[61-63], finance[64], chemistry[65-67], and geosciences[68]. For instance, in geosciences, LLMs have been applied in tasks such as geospatial question answering[69] and named entity recognition[70]. Although there is no specialized evaluation for LLM code generation capabilities, some studies have indirectly quantified their limitations through retrieval-augmented generation (RAG), prompt engineering, or fine-tuning in baseline comparison experiments[5,71]. However, these studies have primarily focused on general-purpose code generation, lacking evaluations specific to geospatial code generation.

## 2.4 Performance Improvement of Large Language Models

With the exploration of LLM capability boundaries, their limitations have gradually emerged[72], and various approaches for improving performance have been developed, primarily categorized into model optimization and usage strategy optimization. Model optimization includes full fine-tuning[73] and incremental fine-tuning[74], while usage strategy optimization encompasses methods such as agent collaboration[75], retrieval-augmented generation[76], prompt engineering[77], few-shot learning[78], Chain of Thought (CoT) design[79], and multi-round voting[37]. The effectiveness of these strategies often depends on available computing resources and task requirements. For highly specialized tasks like geospatial code generation, characterized by rapid corpus updates and long input-output sequences, achieving a balance between optimization cost and performance enhancement remains a current research focus and challenge.

## 3 GeoCode-Bench Construction

### 3.1 Construction Process

Benchmark datasets built on real-world samples can more effectively explore the capability boundaries of LLMs and provide users with a practical evaluation tool. To create a high-quality, comprehensive, and large-scale geospatial code benchmark dataset, the construction process was divided into three phases: "Basic Data Collection—Evaluation Set Construction—Quality Validation". To expand the scope of problem collection, reduce manual work, and minimize subjective bias, this study employed the Self-Instruct framework[80] and rule-based machine learning methods to automatically generate the evaluation set. Self-Instruct framework uses few-shot learning to construct prompts based on reference data and samples, guiding the model to generate questions from real data. Rule-based methods generate questions by slicing or masking after traversing structured data. For instance, cell selection or fill-in-the-blank questions can be generated with tabular data, while code completion questions can be generated via masking code data. After constructing the evaluation set, we manually reviewed the questions to ensure comprehensive coverage, reduce redundancy, and verify answer accuracy, ultimately forming the high-quality

GeoCode-Bench evaluation dataset. The specific process is illustrated in Figure 2.

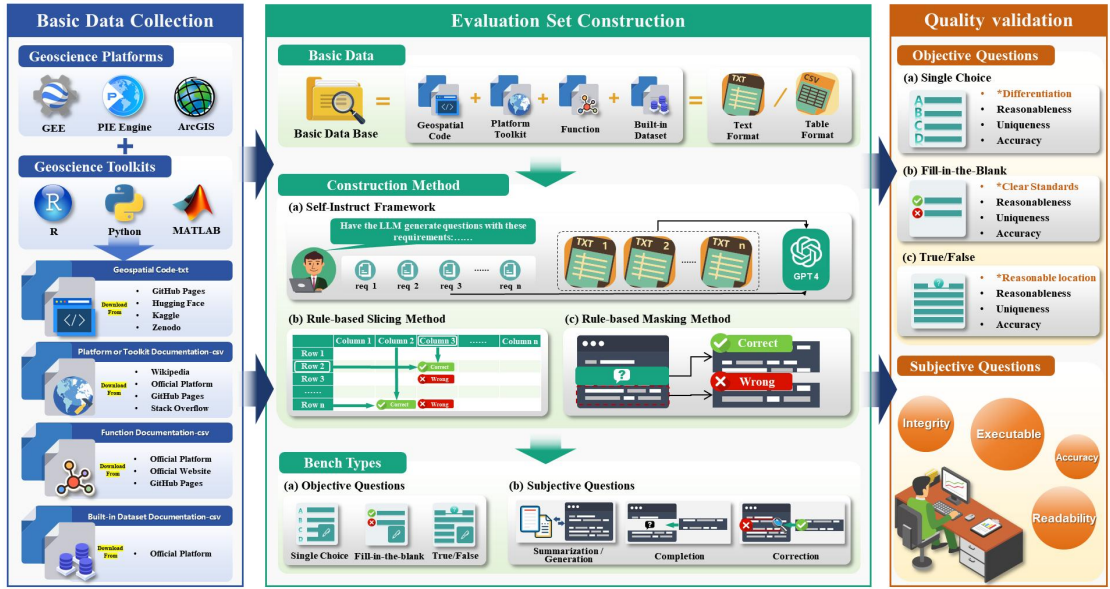

Figure.2 GeoCode-Bench Construction Process Diagram

## 3.2 Basic Data Collection

The GeoCode-Bench dataset is based on four types of data: real code, platform or toolkit introduction, function documentation, and built-in dataset documentation. The real code includes secondarily encapsulated code from various geospatial analysis platforms (e.g., Google Earth Engine, PIE Engine, ArcGIS) based on foundational programming languages or code from specialized geospatial processing toolkits (e.g., Matlab's Mapping Toolbox, R's terra, Python's GDAL and rasterio). The data was sourced from open-source platforms such as GitHub, Hugging Face, Kaggle, and Zenodo, and stored in txt format, totaling 18,286 entries. Platform or toolkit introduction documents were collected from Wikipedia, official platform documentation, GitHub project pages, and technical forums such as Stack Overflow, and stored in txt format, totaling 28 entries. Function documentation for each geospatial analysis platform was sourced from the platform's official documentation, while function documentation for geospatial toolkits came from their official websites and GitHub pages, stored in csv format, totaling 8,729 entries. Both the GEE and PIE Engine platforms have numerous pre-processed remote sensing datasets which users can directly access for analysis or mapping. We collected information on dataset addresses, names, keywords, and content from the official documentation of GEE and PIE Engine cloud platforms, stored in txt format, totaling 2,732 entries. A summary of the data collection is provided in Table 1.

Table.1 Basic Data Collection Overview Table

| Data Type | Platform or Toolkit | Download Source | Storage Format | Quantity (Items) |
|---|---|---|---|---|
| Real Geospatial Code | GEE, PIE Engine, ArcGIS, R, Matlab, Python | GitHub Pages, Hugging Face, Kaggle, Zenodo | txt | 18286 |
| Platform or Toolkit Documentation | GEE, PIE Engine, ArcGIS, R, Matlab, Python | Wikipedia, Official Platform, GitHub Pages, Stack Overflow | txt | 28 |
| Function Documentation | GEE, PIE Engine, ArcGIS, R, Matlab, Python | Official Platform, Official Website, GitHub Pages | csv | 8729 |
| Built-in Dataset Documentation | GEE, PIE Engine | Official Platform | csv | 2732 |

## 3.3 Evaluation Set Construction

The evaluation set comprises seven types of questions: three objective categories—multiple-choice, fill-in-the-blank, and true/false questions—as well as four subjective categories, including summarization, code generation, code completion, and code correction. Objective questions cover a single knowledge dimension with unique, definite answers and are generated based on real information. Multiple-choice questions provide four options; fill-in-the-blank questions contain a single blank; true/false questions involve binary judgments on core concepts. Subjective questions consist of a question, a prompt, and a reference answer. This section introduces the construction methods and rationale for each question type based on the three evaluation dimensions, along with examples.

### 3.3.1 Cognition and Memory

**Geospatial analysis platforms or toolkits knowledge** includes details such as required environments, programming languages, data types, and functionalities[81,82]. This information is used to assess whether LLMs possess the foundational knowledge needed for code generation, enabling them to select the appropriate platform or toolkit according to user requirements. The evaluation is conducted in the form of multiple-choice questions, which are generated by the SELF-INSTRUCT framework based on platform or toolkit introduction documents. The prompt design and example questions are shown in Figure 3.

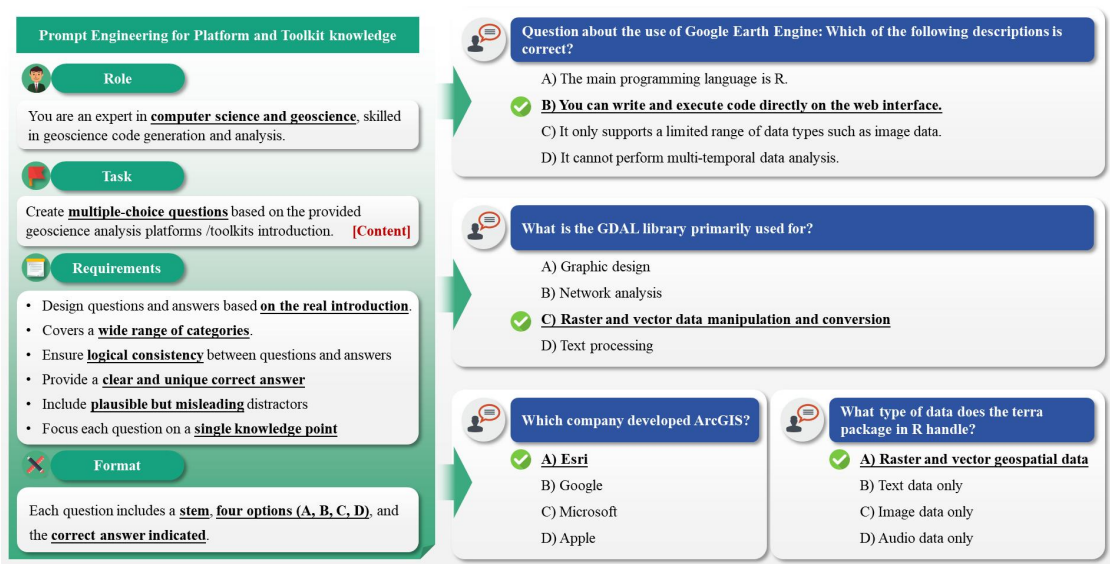

Figure.3 "Platform or Toolkit Knowledge" Cognition Test Construction and Example

**Geospatial function knowledge** involves documentation of platforms or toolkits, including function names, affiliated platforms or toolkits, programming languages, parameters, functionalities, input and output data types, among other details. This knowledge is used to assess LLMs' understanding of syntax of geospatial processing platforms. The evaluation uses both multiple-choice and true/false questions. Correct answers are generated based on slices of function documentation, while incorrect answers are generated through cross-matching. Due to the length and complexity of parameter information, it is unsuitable for use as multiple-choice options; thus, it is evaluated using true/false questions that assess the existence, descriptions, and data types of parameters. Other attributes are assessed through multiple-choice questions. The slice-matching method and example questions are illustrated in Figure 4.

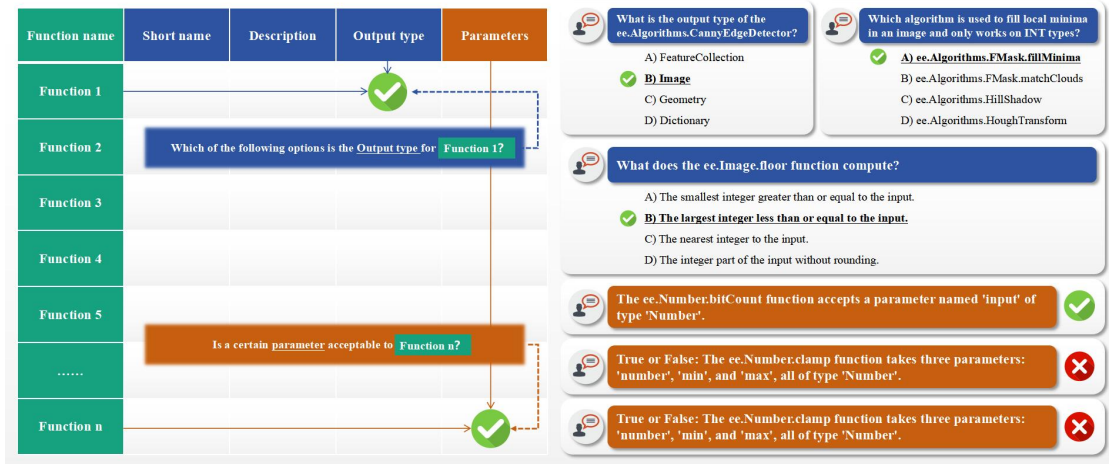

Figure.4 "Function Knowledge" Cognition Test Construction and Example

**Built-in datasets** are stored in hierarchical directories on cloud platforms and include information such as names, paths, platforms, keywords, and spatiotemporal attributes[83]. LLMs need to understand this information to correctly index datasets when generating geospatial code. The evaluation uses both multiple-choice and true/false questions. Correct answers are generated based on slices of built-in dataset documentation, while incorrect answers are generated through cross-matching. Dataset path information is lengthy and thus evaluated through true/false questions, while other attributes are assessed using multiple-choice questions. The slice-matching method and example questions are shown in Figure 5.

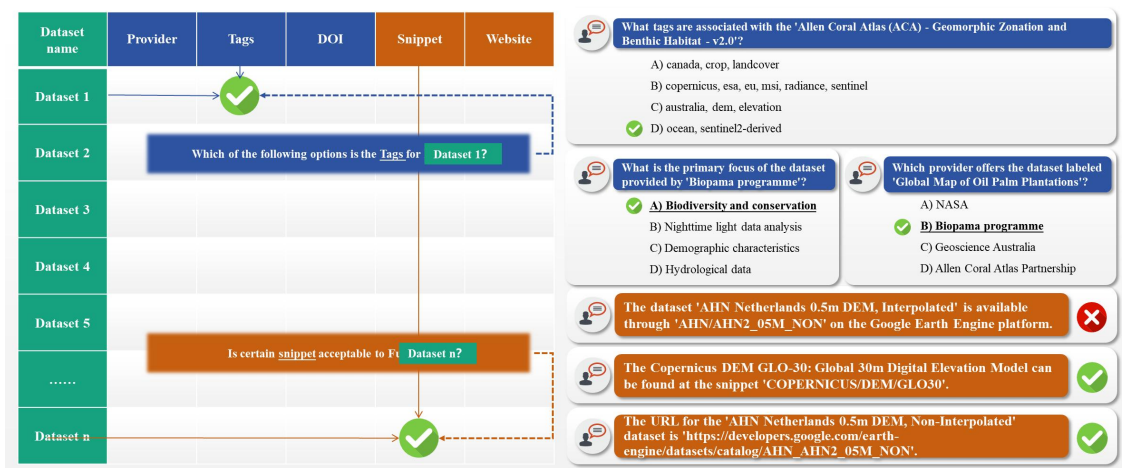

Figure.5 "Built-in Dataset Knowledge" Cognition Test Construction and Example

### 3.3.2 Comprehension and Interpretation

A complete geospatial script typically includes four types of entity information: data source, spatial extent, temporal extent, and functional processes[84]. Accurately identifying this information is crucial for future modifications, corrections, replacements of data sources, or adjustments to functional processes, ensuring that corresponding content is properly retained or updated. These four types of entity information are extracted and recorded in a structured table through prompts guiding LLMs to parse them one by one. The evaluation is conducted using fill-in-the-blank and true/false questions: explicit information such as data source, spatial extent, and temporal extent is assessed through fill-in-the-blank questions to evaluate the model's ability to locate and extract entities, with each question involving only one type of entity. Functional processes, as implicit information, are evaluated using true/false questions to assess the models' semantic understanding capability, including the presence and number of functionalities. The question construction method and examples are illustrated in Figure 6.

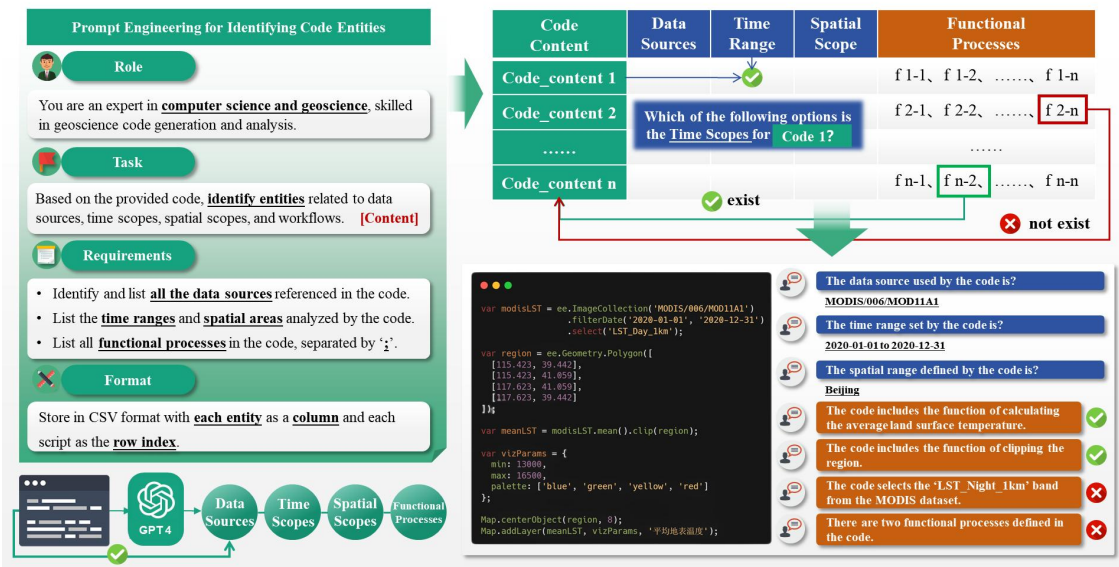

Figure.6 "Entity Recognition" Cognition Test Construction and Example

Before generate code, users often need to interpret published research code or platform code as a reference, learning function usage or experiences, then modifying or recombining code according to their requirements. In this context, LLMs need to generate code summaries[85]. Code summarization by LLMs is the inverse process of code generation, and its core lies in assessing the model's understanding and abstraction abilities for geospatial code. In this paper, we designed prompts and used GPT-4 to generate standard answers, requiring coverage of overall objectives, data sources, temporal extent, spatial extent, output data format, and a summary of functional processes. These questions are subjective, and the specific construction process and examples are shown in Figure 7-(a).

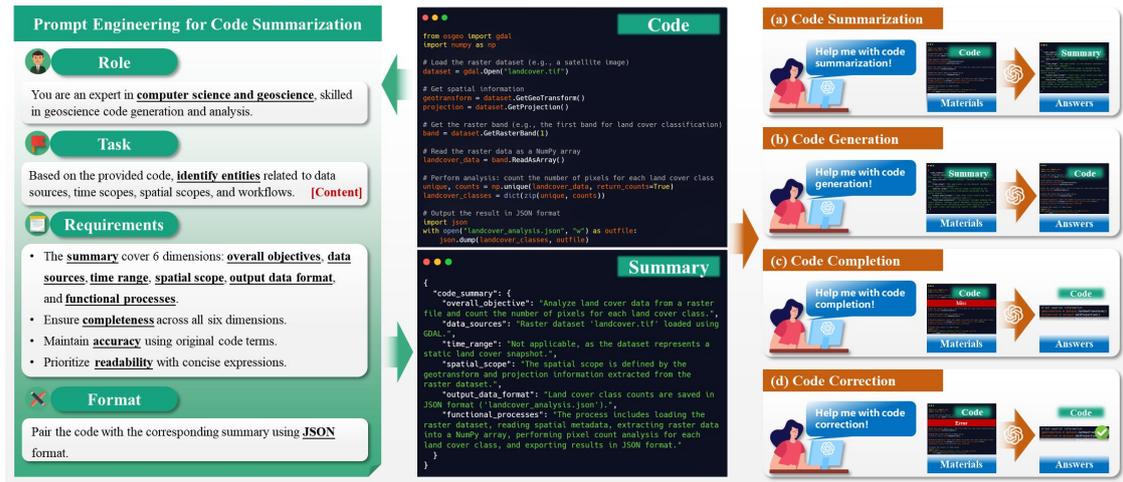

Figure.7 "Subjective Questions" Cognition Test Construction and Example

### 3.3.3 Innovation and Creation

Evaluating a model's ability to generate complete geospatial code according to instructions is a direct way to assess its geospatial coding knowledge. Although code summarization and code generation are inverse tasks involving natural language and programming language, summarization from code has complete information, accurate content and clear processing steps, while generating executable code requires responding to incomplete requirements—usually limited to data sources, temporal extent, spatial extent, and output data types[71]. We designed paired code and summary questions to comparatively evaluate the model's code generation capabilities under different instructions. The code generation capability evaluation question construction method and examples are illustrated in Figure 7-(b).

The code obtained may sometimes be incomplete in three main situations. First,

functional code that lacks partial data or parameter information, requiring the model to locate and complete it according to user instructions. Second, functional code missing some statement, requiring the model to identify errors and complete them. Third, functionally incomplete code, requiring the model to write missing parts of the code according to user instructions[86]. In this paper, we used real code and employed a rule-based masking method to cover data parameters, logical statements, and key code blocks, forming questions where the covered part is the answer, thus constructing the evaluation set. The code completion capability evaluation question construction method and examples are illustrated in Figure 7-(c).

For missing information, LLMs need to locate blank positions and complete them using contextual logic, while for incorrect or unsatisfactory information, the model needs to locate and correct the errors. These two capabilities belong to different evaluation dimensions and need to be evaluated separately. Unlike the code completion task, the code correction evaluation mainly involves three situations: first, replacing part of the data or parameters in the complete code; second, deleting some functional statements; and third, adjusting the order of functional processes or function implementations. Based on these three scenarios and expert experience, we constructed the evaluation set from real code, where the question is the complete code to be modified, and the answer is the modified code[87]. The code correction capability evaluation question construction method and examples are shown in Figure 7-(d).

### 3.4 Expert Correction
All questions must be reviewed and corrected by experts to ensure answer accuracy and conduct multidimensional analysis and validation of each question type. For multiple-choice questions, experts need to confirm the distinction and rationality of options, ensuring answer uniqueness. For true/false questions, evaluation criteria must be clear and unambiguous. For fill-in-the-blank questions, the blank position should be reasonable and have a unique answer. For subjective questions such as summarization, code generation, code completion, and code correction, text questions are manually corrected to ensure completeness, readability, and accuracy, while code questions are also verified for executability. The corrected question types and their quantities are summarized based on the eight capability levels across three dimensions, as shown in Table 2.

Table.2 Evaluation Question Type Data Table

| Dimension | Ability Level | Question Type | Quantity |
|---|---|---|---|
| Cognition and Memory | Platform, Toolkit | Multiple Choice Question | 2000 |
| | Function | Multiple Choice Question | 1500 |
| | | True/False Question | 500 |
| | Built-in Dataset | Multiple Choice Question | 1500 |
| | | True/False Question | 500 |
| Understanding and Interpretation | Entity Recognition | Fill-in-the-Blank Question | 1500 |
| | | True/False Question | 500 |
| | Code Summarization | Subjective Question | 250 |
| Innovation and Creation | Code Generation | Subjective Question | 250 |
| | Code Completion | Subjective Question | 250 |
| | Code Correction | Subjective Question | 250 |

## 4  Evaluation Methodology and Results

### 4.1 Evaluated Models

This study evaluates commercial LLMs, general-purpose LLMs, and general-purpose code generation models, all of which are representative versions that performed well or exhibited notable features in their respective domains at the time of the experiment. These models vary in parameter scale, can be categorized into open-source and closed-source ones. Such comparative design allows us to reveal performance differences across various model properties, providing references for model selection. The models to be evaluated and their relevant information are presented in Table 3.

Table.3 Information Table of Models to be Evaluated

| Model Type | Model Name | Company | Size | Date | Open Source |
|---|---|---|---|---|---|
| Commercial Closed-Source LLMs | GPT-4 | OpenAI | N/A | 2023 | No |
| | Claude-3-Opus | Anthropic | N/A | 2023 | No |
| | ERNIE-4.0 | Baidu | 8B | 2023 | No |
| **General Open-Source LLMs** | LLaMA3 | Meta AI | 8B、70B | 2024 | Yes |
| | PaLM | Google | 540B | 2022 | No |
| | BLOOM | BigScience | 176B | 2022 | Yes |
| **Open-Source Code-Generation LLMs** | CodeGemma | Google | 7B | 2023 | Yes |
| | StarCoder 2 | BigCode | 15B | 2023 | Yes |
| | CodeQwen | Alibaba | 14B | 2023 | Yes |
| | WizardCoder | WizardLM | 15B | 2023 | Yes |
| | Code Llama | Meta AI | 7B、13B、34B | 2023 | Yes |
| | OctoCoder | Hugging Face | 15.5B | 2023 | Yes |
| | CodeGeeX2 | THU/IDEA | 6B | 2023 | Yes |
| | Codex | OpenAI | 12B | 2021 | No |
| | AlphaCode | DeepMind | 1.1B | 2022 | No |
| | CodeT5+ | Salesforce Research | 770M、2B、16B | 2023 | Yes |

## 4.2 Evaluation Scheme

The evaluation follows an "Answer-Judge-Analyze" (AJA) framework.

In the answering phase, we designed specific prompts for each question type to guide the evaluated LLMs in answering the GeoCode-Bench questions. To explore the impact of different prompting strategies on model capabilities, the study also set up control experiments: for multiple-choice and true/false questions, we compared single-round inquiry with five-round inquiry; for fill-in-the-blank questions, we compared zero-shot with few-shot answering; for the four types of subjective questions—summarization, code generation, code completion, and code correction—direct inquiry was compared with step-by-step answering based on Chain of Thought (CoT) reasoning. The answering methods and designs for each question type are shown in Figure 8.

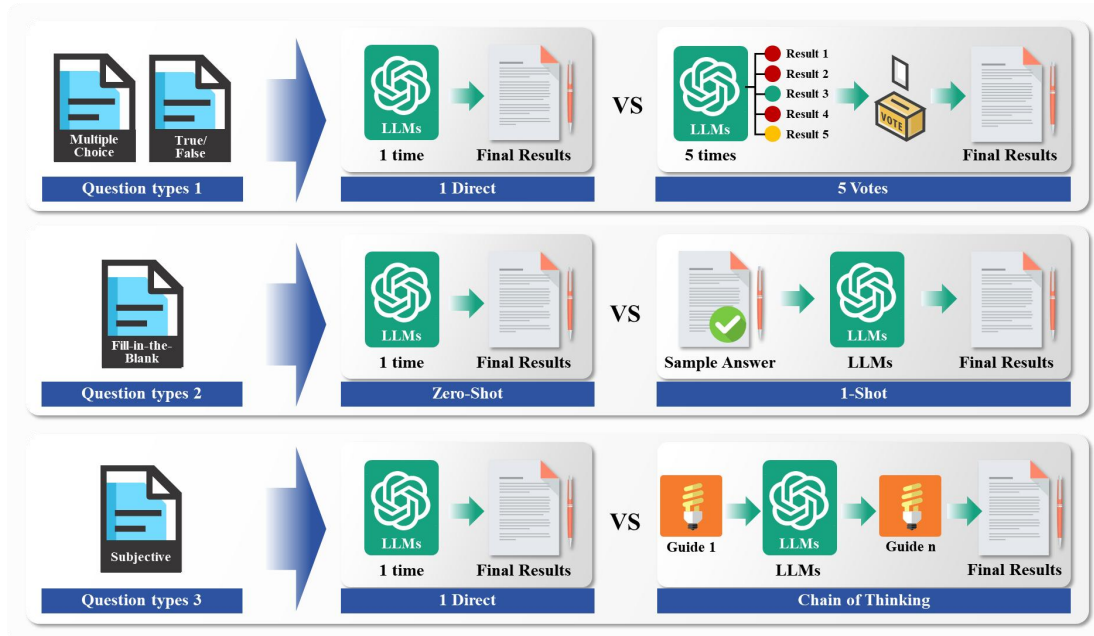

Figure.8 "Answer-Judge-Analyze" Framework Diagram

In the judging phase, explicit evaluation criteria and methods were set for each question type. For objective questions such as multiple-choice and true/false questions, accuracy was calculated by matching the answer with the correct answer. Fill-in-the-blank questions were evaluated using string similarity to assess accuracy. For subjective questions, evaluation criteria for summarization included completeness, accuracy, and textual readability; code generation questions focused on executability, entity accuracy, and code readability. For code completion and code correction, the evaluation criteria were the correctness of the completion location, content accuracy, and executability. The executability of subjective questions was verified by experts on the platform, while readability was assessed by experts using a one-way blind scoring method, wherein answers generated by different models for the same question were randomly sorted and rated by experts on a scale of 1-10. Each answer is scored independently by five experts, and the final score is the average of five scores. Other metrics were scored by GPT-4, with scores ranging from 1-10. The scores were then aggregated to obtain the metric scores for each question type, and the detailed evaluation criteria, explanations, and methods are provided in Table 4.

Table.4 Evaluation Metrics and Methods Table for Each Question Type

| Question Type | Evaluation Standard | Evaluation Method |
|---|---|---|
| **Multiple-choice Question** | Accuracy | Answer matching |
| **True/False Question** | Accuracy | Answer matching |
| **Fill-in-the-blank** | Accuracy | String similarity |
| | Completeness | GPT-4 scoring |
| **Code Summarization** | Accuracy | GPT-4 scoring |
| | Text Readability | Expert scoring |
| | Executability | Expert scoring |
| **Code Generation** | Entity Accuracy | GPT-4 scoring |
| | Code Readability | Expert scoring |
| | Position Accuracy | GPT-4 scoring |
| **Code Completion/Correction** | Content Accuracy | GPT-4 scoring |
| | Executability | Expert scoring |

Finally, an in-depth analysis was conducted based on data from each metric, focusing on the impact of different model choices and prompting strategies on LLM performance across different tasks. The analysis aimed to identify regular, anomalous, unexpected phenomena, and potential innovations, extracting key data to support the evaluation conclusions.

## 4.3 Evaluation Results

### 4.3.1 Platform or Toolkit Knowledge

The evaluation results are shown in Table 5. The Acc@1 values of the models ranged from 0.6420 (CodeT5+-770M) to 0.9435 (GPT-4), indicating that all LLMs reached at least a passing level in the "Platform or Toolkit Knowledge" task, but with significant performance differences. Commercia closed-source l models performed outstandingly, with Acc@1 values exceeding 0.94 for GPT-4 and Claude-3-Opus, and minimal improvement from the voting strategy, indicating high accuracy and stability in their initial responses. This suggests reduced need for repeated interactions when applied in such tasks, demonstrating significant generalizability and stability. While open-source general-purpose models lagged slightly, LLaMA3-8B's Acc@5 improved considerably, even surpassing GPT-4, indicating its potential as an alternative in resource-constrained scenarios. Code generation models (e.g.,

StarCoder 2-15B, Code Llama series) did not perform well in Acc@1, likely due to their training data focusing on programming languages and lacking coverage of geospatial knowledge. The voting strategy improved model performance by 0.4% (CodeT5+-16B) to 4.95% (CodeQwen-14B), validating its effectiveness to some extent.

Table.5 Evaluation Results Table for "Platform or Toolkit Knowledge"

| Category | Model Name | Acc@1 | Acc@5 | Difference |
|---|---|---|---|---|
| Commercial Closed-Source | GPT-4 | 0.9435 | 0.9515 | +0.0080 |
|  | Claude-3-Opus | 0.9405 | 0.9675 | +0.0270 |
|  | ERNIE-4.0 | 0.8860 | 0.8970 | +0.0110 |
|  | **Overall** | **0.9233** | **0.9387** | **+0.0153** |
| General Open-Source | LLaMA3-8B | 0.9115 | 0.9560 | +0.0445 |
|  | LLaMA3-70B | 0.9230 | 0.9410 | +0.0180 |
|  | PaLM-540B | 0.9095 | 0.9455 | +0.0360 |
|  | BLOOM-176B | 0.9135 | 0.9270 | +0.0135 |
|  | **Overall** | **0.9144** | **0.9424** | **+0.0280** |
| Open-Source Code-Generation | CodeGemma-7B | 0.7685 | 0.7740 | +0.0055 |
|  | StarCoder 2-15B | 0.8510 | 0.8830 | +0.0320 |
|  | CodeQwen-14B | 0.8505 | 0.9000 | +0.0495 |
|  | WizardCoder-15B | 0.8550 | 0.8675 | +0.0125 |
|  | Code Llama-7B | 0.8020 | 0.8385 | +0.0365 |
|  | Code Llama-13B | 0.8540 | 0.8730 | +0.0190 |
|  | Code Llama-34B | 0.8405 | 0.8865 | +0.0460 |
|  | OctoCoder-15.5B | 0.8465 | 0.8765 | +0.0300 |
|  | CodeGeeX2-6B | 0.8195 | 0.8345 | +0.0150 |
|  | Codex-12B | 0.7745 | 0.7840 | +0.0095 |
|  | AlphaCode-1.1B | 0.7475 | 0.7880 | +0.0405 |
|  | CodeT5+-770M | 0.6420 | 0.6890 | +0.0470 |
|  | CodeT5+-2B | 0.6930 | 0.7175 | +0.0245 |
|  | CodeT5+-16B | 0.7080 | 0.7120 | +0.0040 |
|  | **Overall** | **0.7895** | **0.8160** | **+0.0265** |

### 4.3.2 Function Knowledge

The evaluation results for Geospatial Function Knowledge tasks are presented in Table 6. The models showed weaker mastery in this task compared to the Platform or Toolkit Knowledge tasks. The Acc@1 range was 0.5190 (CodeT5+-770M) to 0.7435 (GPT-4), with Acc@5 ranging from 0.5320 (CodeT5+-2B) to 0.7495 (GPT-4). The average Acc@1 of closed-source commercial models was 0.6970, significantly outperforming open-source models (0.6204) and code generation models (0.5696). GPT-4 led with 0.7435, demonstrating high initial response accuracy and stability. In contrast, open-source general-purpose models had lower Acc@1 values but showed greater improvement through the voting strategy, such as LLaMA3-8B's improvement of 0.1000. Code generation models showed uneven performance. StarCoder 2-15B (0.6360) and Codex-12B (0.6280) had Acc@1 values close to or exceeding some general-purpose models, while the CodeT5 series performed poorly, falling below the passing level and highlighting the lack of geospatial function knowledge. Overall, the improvement of code generation models was limited, likely due to insufficient geospatial domain coverage in their training data, resulting in non-significant improvement in multi-round responses.

Table.6 Evaluation Results Table for "Function knowledge"

| Category | Model Name | Acc@1 | Acc@5 | Difference |
|---|---|---|---|---|
| **Commercial Closed-Source** | GPT-4 | 0.7435 | 0.7495 | +0.0060 |
| | Claude-3-Opus | 0.6960 | 0.7185 | +0.0225 |
| | ERNIE-4.0 | 0.6515 | 0.6955 | +0.0440 |
| | **Overall** | **0.6970** | **0.7212** | **+0.0242** |
| **General Open-Source** | LLaMA3-8B | 0.5880 | 0.6880 | +0.1000 |
| | LLaMA3-70B | 0.6420 | 0.6780 | +0.0360 |
| | PaLM-540B | 0.6695 | 0.7445 | +0.0750 |
| | BLOOM-176B | 0.5820 | 0.5970 | +0.0150 |
| | **Overall** | **0.6204** | **0.6769** | **+0.0565** |
| **Open-Source Code-Generation** | CodeGemma-7B | 0.5675 | 0.5800 | +0.0125 |
| | StarCoder 2-15B | 0.6360 | 0.6840 | +0.0480 |
| | CodeQwen-14B | 0.6045 | 0.6310 | +0.0265 |
| | WizardCoder-15B | 0.6115 | 0.6340 | +0.0225 |
| | Code Llama-7B | 0.5540 | 0.5940 | +0.0400 |
| | Code Llama-13B | 0.5570 | 0.6025 | +0.0455 |
| | Code Llama-34B | 0.5955 | 0.6105 | +0.0150 |
| | OctoCoder-15.5B | 0.5735 | 0.6120 | +0.0385 |
| | CodeGeeX2-6B | 0.5415 | 0.6015 | +0.0600 |
| | Codex-12B | 0.6280 | 0.6555 | +0.0275 |
| | AlphaCode-1.1B | 0.5355 | 0.5430 | +0.0075 |
| | CodeT5+-770M | 0.5190 | 0.5370 | +0.0180 |
| | CodeT5+-2B | 0.5210 | 0.5320 | +0.0110 |
| | CodeT5+-16B | 0.5295 | 0.5390 | +0.0095 |

| Category | Model Name | Acc@1 | Acc@5 | Difference |
|---|---|---|---|---|
| | Overall | 0.5696 | 0.5969 | +0.0273 |

### 4.3.3 Built-in Dataset Knowledge

The evaluation results are shown in Table 7. Due to the scattered nature of built-in dataset information for GEE and PIE Engine across different pages of official websites, and the limited systematic integration of geospatial datasets into LLM training corpora, all models struggled to grasp their complete structure and hierarchical directories. In this evaluation, Acc@1 values ranged from 0.1170 (CodeT5+-770M) to 0.4325 (GPT-4), with Acc@5 values ranging from 0.1160 to 0.4385, all falling well below the passing threshold. Even the best-performing GPT-4 achieved only 0.4325 for Acc@1. Claude-3-Opus had an Acc@1 of 0.3985, with a slight drop to 0.3960 for Acc@5, while LLaMA3-8B had an Acc@1 of 0.2785, which decreased to 0.2735 for Acc@5, indicating that multiple attempts did not yield improvement, and that the voting strategy struggled to enhance performance in the absence of relevant training data. Code generation models performed particularly poorly, with an average Acc@1 of just 0.2051, even lower than the 25% accuracy rate expected from random guessing, reflecting a significant lack of built-in dataset knowledge.

Table.7 Evaluation Results Table for "Built-in Dataset Knowledge"

| Category | Model Name | Acc@1 | Acc@5 | Difference |
|---|---|---|---|---|
| Commercial Closed-Source | GPT-4 | 0.4325 | 0.4385 | +0.0060 |
| | Claude-3-Opus | 0.3985 | 0.3960 | -0.0025 |
| | ERNIE-4.0 | 0.3840 | 0.3890 | +0.0050 |
| | **Overall** | **0.4050** | **0.4078** | **+0.0028** |
| General Open-Source | LLaMA3-8B | 0.2785 | 0.2735 | -0.0050 |
| | LLaMA3-70B | 0.3675 | 0.3800 | +0.0125 |
| | PaLM-540B | 0.4155 | 0.4305 | +0.0150 |
| | BLOOM-176B | 0.2915 | 0.2950 | +0.0035 |
| | **Overall** | **0.3383** | **0.3448** | **+0.0065** |
| Open-Source Code-Generation | CodeGemma-7B | 0.2015 | 0.1940 | -0.0075 |
| | StarCoder 2-15B | 0.3050 | 0.3140 | +0.0090 |
| | CodeQwen-14B | 0.2650 | 0.2590 | -0.0060 |

| Category | Model Name | Acc@1 | Acc@5 | Difference |
|---|---|---|---|---|
| | WizardCoder-15B | 0.2770 | 0.2790 | +0.0020 |
| | Code Llama-7B | 0.1775 | 0.1735 | -0.0040 |
| | Code Llama-13B | 0.1895 | 0.1995 | +0.0100 |
| | Code Llama-34B | 0.2515 | 0.2595 | +0.0080 |
| | OctoCoder-15.5B | 0.2150 | 0.2175 | +0.0025 |
| | CodeGeeX2-6B | 0.1635 | 0.1585 | -0.0050 |
| | Codex-12B | 0.2905 | 0.2945 | +0.0040 |
| | AlphaCode-1.1B | 0.1515 | 0.1515 | +0.0000 |
| | CodeT5+-770M | 0.1170 | 0.1160 | -0.0010 |
| | CodeT5+-2B | 0.1285 | 0.1330 | +0.0045 |
| | CodeT5+-16B | 0.1390 | 0.1315 | -0.0075 |
| | **Overall** | **0.2051** | **0.2058** | **+0.0006** |

### 4.3.4 Entity Recognition

The evaluation results are presented in Table 8. For the extraction of four types of geospatial script entities (data source, temporal extent, spatial extent, and functional processes), closed-source commercial models performed excellently, with both Acc@1 and Acc@5 values exceeding 0.85. GPT-4 led in data source extraction (Acc@1 of 0.8953) and functional process extraction (Acc@5 of 0.934), while Claude-3-Opus and ERNIE-4.0 were slightly behind but stable. The average Acc@1 of open-source general-purpose models was 0.7763, and the voting strategy improved Acc@5 to 0.848. In functional process extraction, PaLM-540B and LLaMA3-70B achieved Acc@5 values of 0.932 and 0.89, approaching the performance of commercial closed-source models. Code generation models performed weaker, with an average Acc@1 of only 0.6791, and an Acc@5 average of 0.6384 for functional processes. StarCoder 2-15B showed slight improvement in data source extraction, but overall, the performance was limited, especially for the CodeT5 series, which underperformed in this task. Interestingly, commercial closed-source models and general-purpose code generation models were less proficient in functional process extraction than in other entities, possibly due to the subjective nature of the task and the difficulty in defining the number of functions. Open-source general-purpose models excelled in this task, potentially due to their stronger text interpretation capabilities compared to code generation models focused on structural generation.

Table.8 Evaluation Results Table for "Entity Recognition"

| Category | Model Name | Data sources Time range Spatial scope | | | Functional Processes | | |
|---|---|---|---|---|---|---|---|
| | | 0-shot | 1-shot | Difference | Acc@1 | Acc@5 | Difference |
| Commercial Closed-Source | GPT-4 | 0.8953 | 0.9533 | +0.0580 | 0.886 | 0.934 | +0.048 |
| | Claude-3-Opus | 0.8620 | 0.9233 | +0.0613 | 0.842 | 0.862 | +0.02 |
| | ERNIE-4.0 | 0.8127 | 0.8580 | +0.0453 | 0.812 | 0.818 | +0.006 |
| | **Overall** | **0.8567** | **0.9116** | **+0.0549** | **0.847** | **0.871** | **+0.025** |
| General Open-Source | LLaMA3-8B | 0.7260 | 0.7767 | +0.0507 | 0.762 | 0.778 | +0.016 |
| | LLaMA3-70B | 0.7980 | 0.8280 | +0.0300 | 0.8 | 0.89 | +0.09 |
| | PaLM-540B | 0.8427 | 0.8667 | +0.0240 | 0.872 | 0.932 | +0.06 |
| | BLOOM-176B | 0.7387 | 0.8033 | +0.0647 | 0.764 | 0.79 | +0.026 |
| | **Overall** | **0.7763** | **0.8187** | **+0.0423** | **0.799** | **0.848** | **+0.048** |
| Open-Source Code-Generation | CodeGemma-7B | 0.7007 | 0.8187 | +0.1180 | 0.684 | 0.706 | +0.022 |
| | StarCoder 2-15B | 0.7767 | 0.8420 | +0.0653 | 0.71 | 0.756 | +0.046 |
| | CodeQwen-14B | 0.7540 | 0.8067 | +0.0527 | 0.692 | 0.696 | +0.004 |
| | WizardCoder-15B | 0.7627 | 0.8200 | +0.0573 | 0.634 | 0.646 | +0.012 |
| | Code Llama-7B | 0.6727 | 0.7300 | +0.0573 | 0.672 | 0.684 | +0.012 |
| | Code Llama-13B | 0.6847 | 0.7513 | +0.0667 | 0.648 | 0.656 | +0.008 |
| | Code Llama-34B | 0.6620 | 0.6900 | +0.0280 | 0.624 | 0.656 | +0.032 |
| | OctoCoder-15.5B | 0.7480 | 0.8000 | +0.0520 | 0.582 | 0.59 | +0.008 |
| | CodeGeeX2-6B | 0.6400 | 0.6667 | +0.0267 | 0.616 | 0.622 | +0.006 |
| | Codex-12B | 0.6400 | 0.6667 | +0.0267 | 0.558 | 0.572 | +0.014 |
| | AlphaCode-1.1B | 0.6287 | 0.6600 | +0.0313 | 0.564 | 0.586 | +0.022 |
| | CodeT5+-770M | 0.6187 | 0.6567 | +0.0380 | 0.602 | 0.604 | +0.002 |
| | CodeT5+-2B | 0.6100 | 0.6333 | +0.0233 | 0.59 | 0.596 | +0.006 |
| | CodeT5+-16B | 0.6087 | 0.6267 | +0.0180 | 0.558 | 0.568 | +0.01 |
| | **Overall** | **0.6791** | **0.7192** | **+0.0401** | **0.6239** | **0.6384** | **+0.0146** |

### 4.3.5 Code Summarization

The evaluation results are shown in Table 9. In the code summarization task, model performance was assessed in terms of Completeness, Accuracy, and Text Readability. Completeness ranged from 0.272 (CodeT5+-16B) to 0.996 (GPT-4, CoT). GPT-4 achieved 0.924 in direct generation, which improved to 0.996 with CoT, demonstrating that CoT can effectively enhance summary coverage. CodeGemma-7B showed significant improvement from 0.468 to 0.648, even surpassing some general-purpose models (e.g., LLaMA3-70B), indicating that CoT greatly benefits code generation models. For Accuracy, GPT-4 improved from 0.884 to 0.956, indicating enhanced semantic capture with CoT. Code Llama-34B's accuracy improved from 0.676 to 0.776, but CodeT5+-16B showed smaller improvement, reflecting its limited understanding of code summarization. In terms of Text Readability, some models experienced a decline in readability due to increased content complexity from CoT; for instance, Code Llama-34B's readability dropped from 0.696 to 0.648. Nevertheless, GPT-4's readability improved from 0.912 to 0.956, maintaining a high balance. Overall, CoT effectively enhanced Completeness and Accuracy but may negatively impact Text Readability, particularly in code generation models, suggesting a need to balance detailed reasoning with concise text in future model optimization.

Table.9 Evaluation Results Table for "Code Summarization"

| Category | Model Name | Completeness | | | Accuracy | | | Text Readability | | | Overall | | |
|---|---|---|---|---|---|---|---|---|---|---|---|---|---|
| | | Direct | CoT | Difference | Direct | CoT | Difference | Direct | CoT | Difference | Direct | CoT | Difference |
| Commercial Closed-Source | GPT-4 | 0.924 | 0.996 | +0.072 | 0.884 | 0.956 | +0.072 | 0.912 | 0.956 | +0.044 | 0.907 | 0.969 | +0.062 |
| | Claude-3-Opus | 0.872 | 0.928 | +0.056 | 0.848 | 0.94 | +0.092 | 0.864 | 0.816 | -0.048 | 0.861 | 0.895 | +0.034 |
| | ERNIE-4.0 | 0.784 | 0.912 | +0.128 | 0.804 | 0.892 | +0.088 | 0.832 | 0.86 | +0.028 | 0.807 | 0.888 | +0.081 |
| | **Overall** | **0.8600** | **0.9453** | **+0.0853** | **0.845** | **0.929** | **+0.084** | **0.869** | **0.877** | **+0.008** | **0.858** | **0.917** | **+0.059** |
| General Open-Source | LLaMA3-8B | 0.812 | 0.876 | +0.064 | 0.756 | 0.796 | +0.04 | 0.812 | 0.804 | -0.008 | 0.793 | 0.825 | +0.032 |
| | LLaMA3-70B | 0.74 | 0.792 | +0.052 | 0.788 | 0.872 | +0.084 | 0.82 | 0.868 | +0.048 | 0.783 | 0.844 | +0.061 |
| | PaLM-540B | 0.812 | 0.86 | +0.048 | 0.82 | 0.92 | +0.1 | 0.844 | 0.812 | -0.032 | 0.825 | 0.864 | +0.039 |
| | BLOOM-176B | 0.676 | 0.688 | +0.012 | 0.744 | 0.808 | +0.064 | 0.732 | 0.712 | -0.02 | 0.717 | 0.736 | +0.019 |
| | **Overall** | **0.76** | **0.804** | **+0.044** | **0.777** | **0.849** | **+0.072** | **0.802** | **0.799** | **-0.003** | **0.780** | **0.817** | **+0.038** |
| Open-Source Code-Generation | CodeGemma-7B | 0.468 | 0.648 | +0.18 | 0.7 | 0.776 | +0.076 | 0.744 | 0.716 | -0.028 | 0.637 | 0.713 | +0.076 |
| | StarCoder 2-15B | 0.632 | 0.652 | +0.02 | 0.728 | 0.812 | +0.084 | 0.8 | 0.78 | -0.02 | 0.720 | 0.748 | +0.028 |
| | CodeQwen-14B | 0.492 | 0.62 | +0.128 | 0.7 | 0.74 | +0.04 | 0.756 | 0.744 | -0.012 | 0.649 | 0.701 | +0.052 |
| | WizardCoder-15B | 0.6 | 0.68 | +0.08 | 0.692 | 0.788 | +0.096 | 0.776 | 0.776 | 0 | 0.689 | 0.748 | +0.059 |
| | Code Llama-7B | 0.556 | 0.72 | +0.164 | 0.644 | 0.692 | +0.048 | 0.72 | 0.7 | -0.02 | 0.640 | 0.704 | +0.064 |
| | Code Llama-13B | 0.52 | 0.664 | +0.144 | 0.656 | 0.724 | +0.068 | 0.716 | 0.732 | +0.016 | 0.631 | 0.707 | +0.076 |
| | Code Llama-34B | 0.44 | 0.612 | +0.172 | 0.676 | 0.776 | +0.1 | 0.696 | 0.648 | -0.048 | 0.604 | 0.679 | +0.075 |
| | OctoCoder-15.5B | 0.412 | 0.64 | +0.228 | 0.636 | 0.68 | +0.044 | 0.68 | 0.684 | +0.004 | 0.576 | 0.668 | +0.092 |
| | CodeGeeX2-6B | 0.384 | 0.616 | +0.232 | 0.584 | 0.628 | +0.044 | 0.656 | 0.628 | -0.028 | 0.541 | 0.624 | +0.083 |
| | Codex-12B | 0.344 | 0.592 | +0.248 | 0.6 | 0.672 | +0.072 | 0.632 | 0.652 | +0.02 | 0.525 | 0.639 | +0.114 |

| Category | Model Name | Completeness | | | Accuracy | | | Text Readability | | | Overall | | |
|---|---|---|---|---|---|---|---|---|---|---|---|---|---|
| | | Direct | CoT | Difference | Direct | CoT | Difference | Direct | CoT | Difference | Direct | CoT | Difference |
| | AlphaCode-1.1B | 0.32 | 0.696 | +0.376 | 0.564 | 0.612 | +0.048 | 0.612 | 0.576 | -0.036 | 0.499 | 0.628 | +0.129 |
| | CodeT5+-770M | 0.296 | 0.636 | +0.34 | 0.52 | 0.56 | +0.04 | 0.584 | 0.612 | +0.028 | 0.467 | 0.603 | +0.136 |
| | CodeT5+-2B | 0.28 | 0.7 | +0.42 | 0.54 | 0.62 | +0.08 | 0.568 | 0.56 | -0.008 | 0.463 | 0.627 | +0.164 |
| | CodeT5+-16B | 0.272 | 0.6 | +0.328 | 0.5 | 0.564 | +0.064 | 0.548 | 0.52 | -0.028 | 0.440 | 0.561 | +0.121 |
| | **Overall** | **0.4297** | **0.6483** | **+0.2186** | **0.624** | **0.689** | **+0.065** | **0.678** | **0.666** | **-0.011** | **0.577** | **0.668** | **+0.091** |

### 4.3.6 Code Generation

The evaluation results are shown in Table 10. In the code generation task, model performance focused on Executability, Entity Accuracy, and Code Readability. Overall, code executability was generally low, with scores ranging from 0.208 (CodeT5+-770M, Direct) to 0.584 (GPT-4, CoT). Even Claude-3-Opus, under CoT, only achieved an executability score of 0.58, an improvement of 0.116 but still below the passing threshold. Open-source models performed worse, with most scores below 0.4. Although CoT improved performance, the effect was limited, mainly due to the models' lack of knowledge regarding the function syntax and built-in datasets of geospatial platforms. Even when entities were correctly interpreted in the instructions, models struggled to combine them with the correct syntax, resulting in structurally correct but non-executable code. GPT-4 and Claude-3-Opus performed well in code readability, with limited improvement from CoT. In some cases, such as Claude-3-Opus and StarCoder 2-15B, readability slightly declined, indicating that excessive reasoning could lead to verbose code, compromising conciseness. Although GPT-4's readability was close to perfect (0.964), its executability was only 0.584, showing that fluency and structural clarity do not necessarily translate into executability. Notably, Code Llama-34B showed significant improvement in executability under CoT, increasing from 0.372 to 0.488, despite relatively lower readability (0.72). This reflects a focus on generating executable code during step-by-step reasoning rather than prioritizing syntax simplicity, highlighting the trade-off between executively and readability in increasing the models' performance.

Table.10 Evaluation Results Table for "Code Generation"

| Category | Model Name | Executability | | | Entity Accuracy | | | Code Readability | | | Overall | | |
|---|---|---|---|---|---|---|---|---|---|---|---|---|---|
| | | Direct | CoT | Difference | Direct | CoT | Difference | Direct | CoT | Difference | Direct | CoT | Difference |
| Commercial Closed-Source | GPT-4 | 0.488 | 0.584 | +0.096 | 0.556 | 0.644 | +0.088 | 0.956 | 0.964 | +0.008 | 0.667 | 0.731 | +0.064 |
| | Claude-3-Opus | 0.464 | 0.58 | +0.116 | 0.404 | 0.54 | +0.136 | 0.936 | 0.952 | +0.016 | 0.601 | 0.691 | +0.089 |
| | ERNIE-4.0 | 0.4 | 0.46 | +0.06 | 0.332 | 0.432 | +0.1 | 0.86 | 0.916 | +0.056 | 0.531 | 0.603 | +0.072 |
| | **Overall** | **0.451** | **0.541** | **+0.091** | **0.431** | **0.539** | **+0.108** | **0.917** | **0.944** | **+0.027** | **0.600** | **0.675** | **+0.075** |
| General Open-Source | LLaMA3-8B | 0.244 | 0.324 | +0.08 | 0.312 | 0.408 | +0.096 | 0.768 | 0.712 | -0.056 | 0.441 | 0.481 | +0.040 |
| | LLaMA3-70B | 0.388 | 0.488 | +0.1 | 0.464 | 0.592 | +0.128 | 0.84 | 0.892 | +0.052 | 0.564 | 0.657 | +0.093 |
| | PaLM-540B | 0.272 | 0.376 | +0.104 | 0.204 | 0.344 | +0.14 | 0.876 | 0.952 | +0.076 | 0.451 | 0.557 | +0.107 |
| | BLOOM-176B | 0.256 | 0.308 | +0.052 | 0.316 | 0.428 | +0.112 | 0.788 | 0.828 | +0.04 | 0.453 | 0.521 | +0.068 |
| | **Overall** | **0.290** | **0.374** | **+0.084** | **0.324** | **0.443** | **+0.119** | **0.818** | **0.846** | **+0.028** | **0.477** | **0.554** | **+0.077** |
| Open-Source Code-Generation | CodeGemma-7B | 0.22 | 0.272 | +0.052 | 0.272 | 0.384 | +0.112 | 0.704 | 0.66 | -0.044 | 0.399 | 0.439 | +0.040 |
| | StarCoder 2-15B | 0.44 | 0.552 | +0.112 | 0.5 | 0.632 | +0.132 | 0.916 | 0.872 | -0.044 | 0.619 | 0.685 | +0.067 |
| | CodeQwen-14B | 0.332 | 0.452 | +0.12 | 0.376 | 0.504 | +0.128 | 0.804 | 0.872 | +0.068 | 0.504 | 0.609 | +0.105 |
| | WizardCoder-15B | 0.416 | 0.5 | +0.084 | 0.476 | 0.604 | +0.128 | 0.824 | 0.868 | +0.044 | 0.572 | 0.657 | +0.085 |
| | Code Llama-7B | 0.32 | 0.436 | +0.116 | 0.248 | 0.408 | +0.16 | 0.752 | 0.696 | -0.056 | 0.440 | 0.513 | +0.073 |
| | Code Llama-13B | 0.36 | 0.404 | +0.044 | 0.432 | 0.56 | +0.128 | 0.74 | 0.704 | -0.036 | 0.511 | 0.556 | +0.045 |
| | Code Llama-34B | 0.372 | 0.488 | +0.116 | 0.3 | 0.456 | +0.156 | 0.72 | 0.792 | +0.072 | 0.464 | 0.579 | +0.115 |
| | OctoCoder-15.5B | 0.348 | 0.416 | +0.068 | 0.352 | 0.512 | +0.16 | 0.688 | 0.76 | +0.072 | 0.463 | 0.563 | +0.100 |
| | CodeGeeX2-6B | 0.308 | 0.372 | +0.064 | 0.38 | 0.54 | +0.16 | 0.672 | 0.604 | -0.068 | 0.453 | 0.505 | +0.052 |

| Category | Model Name | Executability | | | Entity Accuracy | | | Code Readability | | | Overall | | |
|---|---|---|---|---|---|---|---|---|---|---|---|---|---|
| | | Direct | CoT | Difference | Direct | CoT | Difference | Direct | CoT | Difference | Direct | CoT | Difference |
| | Codex-12B | 0.424 | 0.524 | +0.1 | 0.468 | 0.628 | +0.16 | 0.892 | 0.936 | +0.044 | 0.595 | 0.696 | +0.101 |
| | AlphaCode-1.1B | 0.232 | 0.34 | +0.108 | 0.172 | 0.252 | +0.08 | 0.652 | 0.712 | +0.06 | 0.352 | 0.435 | +0.083 |
| | CodeT5+-770M | 0.208 | 0.296 | +0.088 | 0.224 | 0.404 | +0.18 | 0.588 | 0.624 | +0.036 | 0.340 | 0.441 | +0.101 |
| | CodeT5+-2B | 0.284 | 0.332 | +0.048 | 0.284 | 0.5 | +0.216 | 0.632 | 0.684 | +0.052 | 0.400 | 0.505 | +0.105 |
| | CodeT5+-16B | 0.296 | 0.376 | +0.08 | 0.34 | 0.384 | +0.044 | 0.632 | 0.676 | +0.044 | 0.423 | 0.479 | +0.056 |
| | **Overall** | **0.326** | **0.411** | **+0.086** | **0.345** | **0.483** | **+0.139** | **0.730** | **0.747** | **+0.017** | **0.467** | **0.547** | **+0.081** |

### 4.3.7 Code Completion

The evaluation results are shown in Table 11. In the code completion task, models performed well in terms of Position Accuracy, particularly GPT-4, Claude-3-Opus, and StarCoder 2-15B, which achieved Position Accuracy scores of 0.808, 0.796, and 0.780 respectively with the CoT strategy. This indicates that the models could effectively identify missing parts of the code and accurately locate areas needing completion. However, Content Accuracy was generally mediocre. Although GPT-4 and PaLM-540B improved their Content Accuracy to 0.624 and 0.612, respectively, with the CoT strategy, overall scores were still not high. Most models showed low Content Accuracy, with the average improving from 0.323 to 0.445, indicating lack of knowledge in syntax when generating completion code. Open-source code generation models like CodeT5+-770M and CodeGemma-7B did not show significant improvement in Content Accuracy even with CoT, scoring only 0.376 and 0.412, respectively. Consequently, models exhibited poor performance in Executability, with even GPT-4 only improving to 0.512 under CoT, while the average score increased from 0.245 to 0.314. Some models, such as AlphaCode-1.1B and CodeT5+-16B, even showed a decline in Executability, further indicating that syntax and functional logic errors in content generation directly affected code executability.

Table.11 Evaluation Results Table for "Code Completion"

| Category | Model Name | Position Accuracy | | | Content Accuracy | | | Executability | | | Overall | | |
|---|---|---|---|---|---|---|---|---|---|---|---|---|---|
| | | Direct | CoT | Difference | Direct | CoT | Difference | Direct | CoT | Difference | Direct | CoT | Difference |
| | GPT-4 | 0.744 | 0.808 | +0.064 | 0.54 | 0.624 | +0.084 | 0.428 | 0.512 | +0.084 | 0.571 | 0.648 | +0.077 |
| Commercial Closed-Source | Claude-3-Opus | 0.732 | 0.796 | +0.064 | 0.48 | 0.56 | +0.08 | 0.384 | 0.444 | +0.06 | 0.532 | 0.600 | +0.068 |
| | ERNIE-4.0 | 0.656 | 0.728 | +0.072 | 0.412 | 0.552 | +0.14 | 0.320 | 0.352 | +0.032 | 0.463 | 0.544 | +0.081 |
| | **Overall** | **0.711** | **0.777** | **+0.067** | **0.477** | **0.579** | **+0.101** | **0.377** | **0.436** | **+0.059** | **0.522** | **0.597** | **+0.076** |
| | LLaMA3-8B | 0.596 | 0.668 | +0.072 | 0.364 | 0.54 | +0.176 | 0.288 | 0.328 | +0.04 | 0.416 | 0.512 | +0.096 |
| | LLaMA3-70B | 0.640 | 0.708 | +0.068 | 0.4 | 0.548 | +0.148 | 0.312 | 0.392 | +0.08 | 0.451 | 0.549 | +0.099 |
| General Open-Source | PaLM-540B | 0.672 | 0.732 | +0.060 | 0.44 | 0.612 | +0.172 | 0.340 | 0.428 | +0.088 | 0.484 | 0.591 | +0.107 |
| | BLOOM-176B | 0.624 | 0.680 | +0.056 | 0.384 | 0.508 | +0.124 | 0.304 | 0.38 | +0.076 | 0.437 | 0.523 | +0.085 |
| | **Overall** | **0.633** | **0.697** | **+0.064** | **0.397** | **0.552** | **+0.155** | **0.311** | **0.382** | **+0.071** | **0.447** | **0.544** | **+0.097** |
| | CodeGemma-7B | 0.520 | 0.580 | +0.060 | 0.288 | 0.412 | +0.124 | 0.212 | 0.296 | +0.084 | 0.340 | 0.429 | +0.089 |
| | StarCoder 2-15B | 0.716 | 0.780 | +0.064 | 0.464 | 0.584 | +0.12 | 0.372 | 0.468 | +0.096 | 0.517 | 0.611 | +0.093 |
| | CodeQwen-14B | 0.612 | 0.660 | +0.048 | 0.352 | 0.496 | +0.144 | 0.268 | 0.3 | +0.032 | 0.411 | 0.485 | +0.075 |
| | WizardCoder-15B | 0.688 | 0.756 | +0.068 | 0.428 | 0.548 | +0.12 | 0.324 | 0.416 | +0.092 | 0.480 | 0.573 | +0.093 |
| Open-Source Code-Generation | Code Llama-7B | 0.548 | 0.584 | +0.036 | 0.316 | 0.444 | +0.128 | 0.244 | 0.236 | -0.008 | 0.369 | 0.421 | +0.052 |
| | Code Llama-13B | 0.564 | 0.620 | +0.056 | 0.328 | 0.412 | +0.084 | 0.248 | 0.316 | +0.068 | 0.380 | 0.449 | +0.069 |
| | Code Llama-34B | 0.580 | 0.636 | +0.056 | 0.34 | 0.496 | +0.156 | 0.252 | 0.332 | +0.08 | 0.391 | 0.488 | +0.097 |
| | OctoCoder-15.5B | 0.536 | 0.592 | +0.056 | 0.304 | 0.432 | +0.128 | 0.236 | 0.312 | +0.076 | 0.359 | 0.445 | +0.087 |
| | CodeGeeX2-6B | 0.504 | 0.544 | +0.040 | 0.276 | 0.376 | +0.1 | 0.204 | 0.276 | +0.072 | 0.328 | 0.399 | +0.071 |
| | Codex-12B | 0.700 | 0.764 | +0.064 | 0.452 | 0.548 | +0.096 | 0.348 | 0.436 | +0.088 | 0.500 | 0.583 | +0.083 |

| Category | Model Name | Position Accuracy | | | Content Accuracy | | | Executability | | | Overall | | |
|---|---|---|---|---|---|---|---|---|---|---|---|---|---|
| | | Direct | CoT | Difference | Direct | CoT | Difference | Direct | CoT | Difference | Direct | CoT | Difference |
| | AlphaCode-1.1B | 0.492 | 0.464 | -0.028 | 0.264 | 0.408 | +0.144 | 0.200 | 0.264 | +0.064 | 0.319 | 0.379 | +0.060 |
| | CodeT5+-770M | 0.444 | 0.480 | +0.036 | 0.224 | 0.376 | +0.152 | 0.160 | 0.22 | +0.06 | 0.276 | 0.359 | +0.083 |
| | CodeT5+-2B | 0.460 | 0.492 | +0.032 | 0.236 | 0.316 | +0.08 | 0.168 | 0.264 | +0.096 | 0.288 | 0.357 | +0.069 |
| | CodeT5+-16B | 0.476 | 0.464 | -0.012 | 0.252 | 0.388 | +0.136 | 0.192 | 0.26 | +0.068 | 0.307 | 0.371 | +0.064 |
| | **Overall** | **0.560** | **0.601** | **+0.041** | **0.323** | **0.445** | **+0.122** | **0.245** | **0.314** | **+0.069** | **0.376** | **0.454** | **+0.078** |

**4.3.8 Code Correction**

The evaluation results are shown in Table 12. Although both code correction and code completion tasks aim to transform defective code into complete executable code, scores for code correction were generally lower than for code completion, with almost all models failing to reach the 0.6 passing threshold. Specifically, improvement in Position Accuracy was very limited in the correction task, with the overall score decreasing slightly from 0.427 to 0.422. Some models, such as CodeGemma-7B and CodeGeeX2-6B, even showed a decline under the CoT strategy, indicating that models struggled to accurately locate errors in logically complete code. Content Accuracy followed a similar trend, with scores for CodeT5+-770M and CodeGeeX2-6B decreasing from 0.376 (in the completion task) to 0.268 and 0.260, respectively, in the correction task, reflecting weaker logical reasoning abilities in the correction task compared to the completion task. As a result, the corrected code still contained syntax or logical errors, making it difficult to execute. For instance, Executability scores for CodeT5+-770M dropped from 0.192 to 0.164, and for AlphaCode-1.1B from 0.264 to 0.140. This could be because the missing parts in the completion task were more apparent, making them easier for models to identify and fill, while hidden errors in the correction task were harder to detect, and models lacked the necessary geospatial syntax knowledge for effective correction.

Table.12 Evaluation Results Table for "Code Correction"

| Category | Model Name | Position Accuracy | | | Content Accuracy | | | Executability | | | Overall | | |
|---|---|---|---|---|---|---|---|---|---|---|---|---|---|
| | | Direct | CoT | Difference | Direct | CoT | Difference | Direct | CoT | Difference | Direct | CoT | Difference |
| Commercial Closed-Source | GPT-4 | 0.600 | 0.712 | +0.112 | 0.452 | 0.524 | +0.072 | 0.388 | 0.412 | +0.024 | 0.480 | 0.549 | +0.069 |
| | Claude-3-Opus | 0.584 | 0.632 | +0.048 | 0.432 | 0.468 | +0.036 | 0.368 | 0.392 | +0.024 | 0.461 | 0.497 | +0.036 |
| | ERNIE-4.0 | 0.512 | 0.528 | +0.016 | 0.332 | 0.404 | +0.072 | 0.292 | 0.320 | +0.028 | 0.379 | 0.417 | +0.039 |
| | **Overall** | **0.565** | **0.624** | **+0.059** | **0.405** | **0.465** | **+0.060** | **0.349** | **0.375** | **+0.025** | **0.440** | **0.488** | **+0.048** |
| General Open-Source | LLaMA3-8B | 0.444 | 0.444 | +0.000 | 0.260 | 0.308 | +0.048 | 0.208 | 0.240 | +0.032 | 0.304 | 0.331 | +0.027 |
| | LLaMA3-70B | 0.496 | 0.504 | +0.008 | 0.352 | 0.424 | +0.072 | 0.272 | 0.300 | +0.028 | 0.373 | 0.409 | +0.036 |
| | PaLM-540B | 0.524 | 0.512 | -0.012 | 0.360 | 0.432 | +0.072 | 0.276 | 0.312 | +0.036 | 0.387 | 0.419 | +0.032 |
| | BLOOM-176B | 0.484 | 0.516 | +0.032 | 0.332 | 0.384 | +0.052 | 0.252 | 0.280 | +0.028 | 0.356 | 0.393 | +0.037 |
| | **Overall** | **0.487** | **0.494** | **+0.007** | **0.326** | **0.387** | **+0.061** | **0.252** | **0.283** | **+0.031** | **0.355** | **0.388** | **+0.033** |
| Open-Source Code-Generation | CodeGemma-7B | 0.392 | 0.376 | -0.016 | 0.228 | 0.268 | +0.040 | 0.184 | 0.212 | +0.028 | 0.268 | 0.285 | +0.017 |
| | StarCoder 2-15B | 0.564 | 0.556 | -0.008 | 0.368 | 0.412 | +0.044 | 0.328 | 0.360 | +0.032 | 0.420 | 0.443 | +0.023 |
| | CodeQwen-14B | 0.432 | 0.436 | +0.004 | 0.284 | 0.320 | +0.036 | 0.232 | 0.264 | +0.032 | 0.316 | 0.340 | +0.024 |
| | WizardCoder-15B | 0.540 | 0.552 | +0.012 | 0.340 | 0.400 | +0.060 | 0.280 | 0.316 | +0.036 | 0.387 | 0.423 | +0.036 |
| | Code Llama-7B | 0.404 | 0.424 | +0.020 | 0.264 | 0.320 | +0.056 | 0.192 | 0.224 | +0.032 | 0.287 | 0.323 | +0.036 |
| | Code Llama-13B | 0.456 | 0.440 | -0.016 | 0.240 | 0.300 | +0.060 | 0.200 | 0.236 | +0.036 | 0.299 | 0.325 | +0.027 |
| | Code Llama-34B | 0.468 | 0.472 | +0.004 | 0.288 | 0.348 | +0.060 | 0.236 | 0.264 | +0.028 | 0.331 | 0.361 | +0.031 |
| | OctoCoder-15.5B | 0.416 | 0.388 | -0.028 | 0.272 | 0.320 | +0.048 | 0.220 | 0.244 | +0.024 | 0.303 | 0.317 | +0.015 |
| | CodeGeeX2-6B | 0.380 | 0.352 | -0.028 | 0.188 | 0.260 | +0.072 | 0.140 | 0.164 | +0.024 | 0.236 | 0.259 | +0.023 |
| | Codex-12B | 0.552 | 0.552 | 0.000 | 0.388 | 0.452 | +0.064 | 0.352 | 0.384 | +0.032 | 0.431 | 0.463 | +0.032 |

| Category | Model Name | Position Accuracy | | | Content Accuracy | | | Executability | | | Overall | | |
|---|---|---|---|---|---|---|---|---|---|---|---|---|---|
| | | Direct | CoT | Difference | Direct | CoT | Difference | Direct | CoT | Difference | Direct | CoT | Difference |
| | AlphaCode-1.1B | 0.364 | 0.356 | -0.008 | 0.184 | 0.224 | +0.040 | 0.104 | 0.140 | +0.036 | 0.217 | 0.240 | +0.023 |
| | CodeT5+-770M | 0.324 | 0.300 | -0.024 | 0.204 | 0.268 | +0.064 | 0.164 | 0.192 | +0.028 | 0.231 | 0.253 | +0.023 |
| | CodeT5+-2B | 0.340 | 0.340 | 0.000 | 0.160 | 0.220 | +0.060 | 0.128 | 0.152 | +0.024 | 0.209 | 0.237 | +0.028 |
| | CodeT5+-16B | 0.352 | 0.360 | +0.008 | 0.152 | 0.192 | +0.040 | 0.108 | 0.140 | +0.032 | 0.204 | 0.231 | +0.027 |
| | **Overall** | **0.427** | **0.422** | **-0.006** | **0.254** | **0.307** | **+0.053** | **0.205** | **0.235** | **+0.030** | **0.296** | **0.321** | **+0.026** |

**4.3.9 Overall Results**

The performance of the 21 models across the three dimensions of "Cognition and Memory—Comprehension and Interpretation—Innovation and Creation" and eight capability levels is illustrated in Figure 9. Overall, none of the models achieved a fully balanced radar chart, indicating that even the most advanced models currently exhibit uneven performance across different capability levels. The envelope of the radar charts shrinks from top to bottom, further revealing that commercial models generally outperform general-purpose LLMs. The relatively weaker performance of code generation models may be attributed to their training, which primarily relies on general code corpora. This has led to a further dilution of geospatial knowledge within the training data, thereby diminishing the models' ability to generate geospatial code. Additionally, each radar chart includes average scores for the eight capability levels, as well as scores after applying improvement strategies. Although some models showed declines in certain capability levels, the majority exhibited significant improvements in average scores and most dimensions.

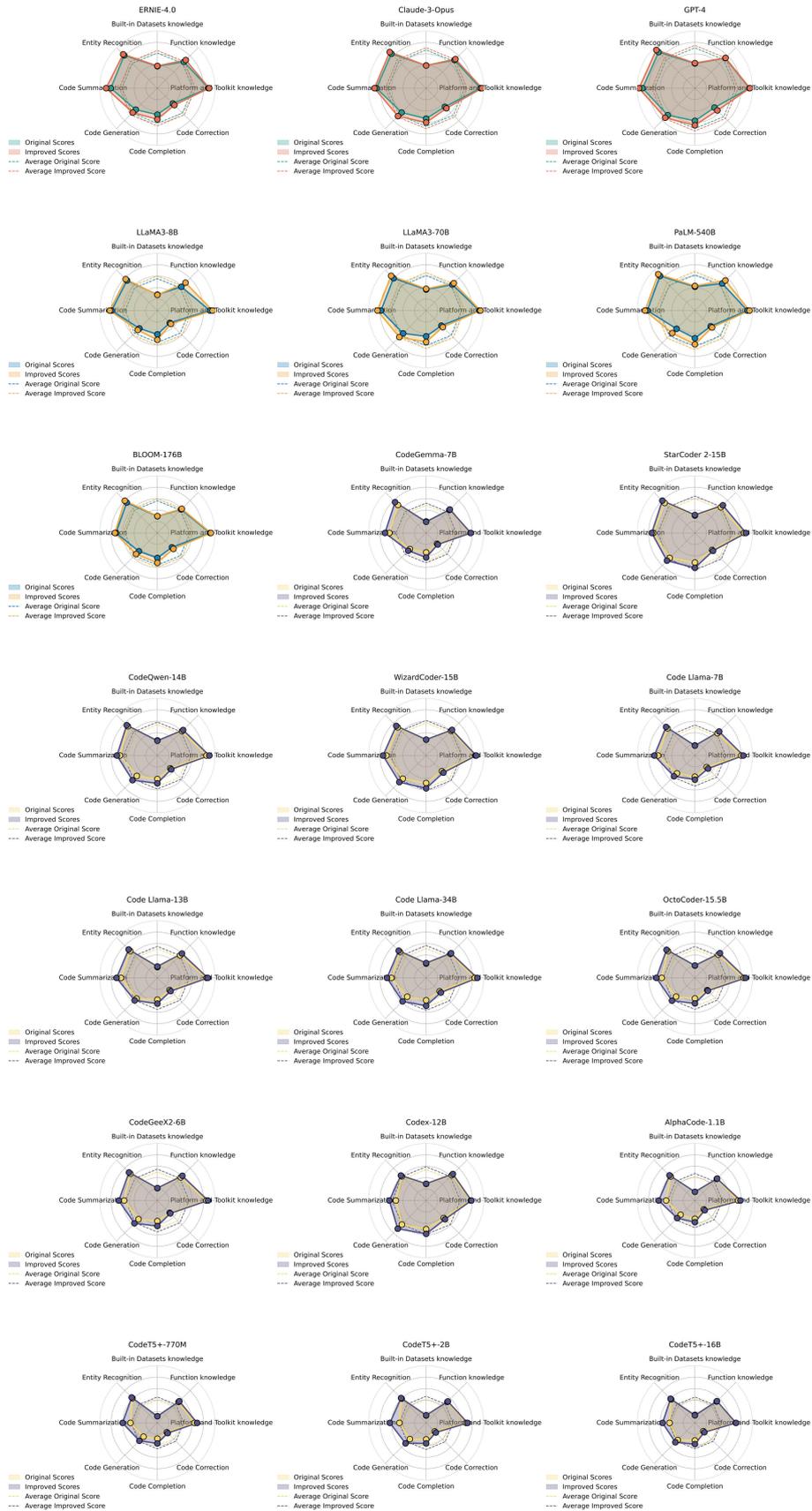

Figure.9 Radar Chart of Model Performance

Comparison of different parameter scales within the same generation of models, as shown in Figure 10, revealed a positive correlation between parameter size and performance for general-purpose open-source LLMs. However, this correlation was not apparent in code generation models, suggesting that parameter size alone does not effectively reflect a model's actual capabilities in the absence of targeted training corpora.

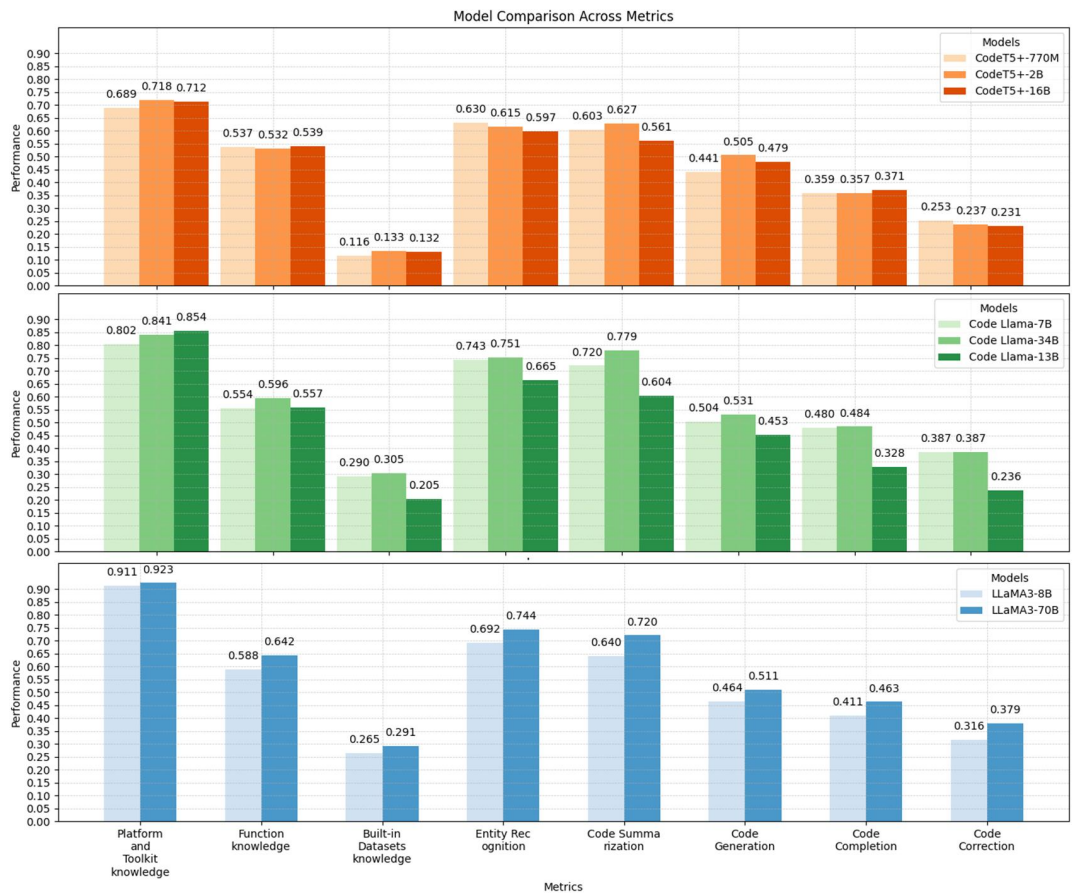

Figure.10 Parameter scale Comparison Bar Chart

A boxplot visualization and analysis of the original and improved metrics across different dimensions for each model, as shown in Figure 11, revealed that models performed best in "Platform and Toolkit Knowledge", "Entity Recognition", and "Code Summarization". Conversely, performance was weaker in "Built-in Dataset Knowledge", "Code Completion", and "Code Correction". Performance in "Function Knowledge" and "Code Generation" fell between the two groups, exhibiting moderate levels.

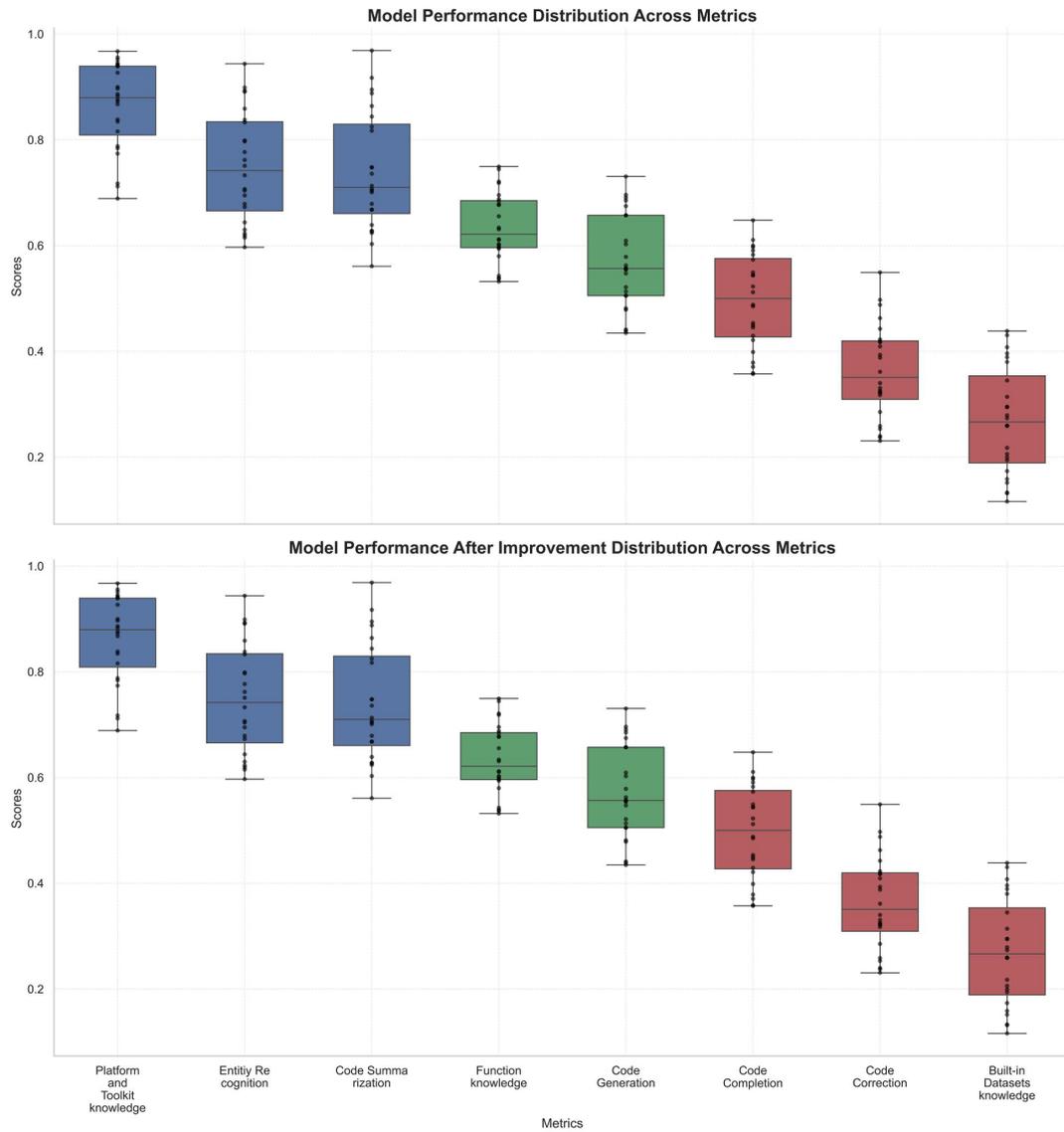

Figure.11 Boxplot Map of Original vs Improved Metrics by Model

## 5 Fine-tuning Case Study

To enhance the performance of large language models (LLMs) in specialized domains, a two-stage optimization approach can be employed: first, pre-training the model using unlabeled domain-specific corpora, followed by instruction fine-tuning to improve task comprehension[88]. As Google Earth Engine (GEE) is one of the biggest and most popular geospatial computing platforms, with numerous user scripts, various geospatial operators and built-in datasets and clear documents, the study focused on model's capability on Google Earth Engine. We collected GEE JavaScript scripts and constructed a set of pre-training corpus GJCode-PT, instruction data GJCode-SFT, and evaluation data GJCode-Eval based on the initial foundational data and the GeoCode-Bench dataset. The general-purpose code generation model Code LLaMA-7B was selected for fine-tuning using these constructed datasets. The performance of the fine-tuned model was subsequently validated on the evaluation set. The entire process included corpus preparation, pre-training, instruction fine-tuning, and evaluation testing, aiming to explore how this staged training process enhances the model's specialized capabilities in geospatial code generation tasks. The detailed process is illustrated in Figure 12.

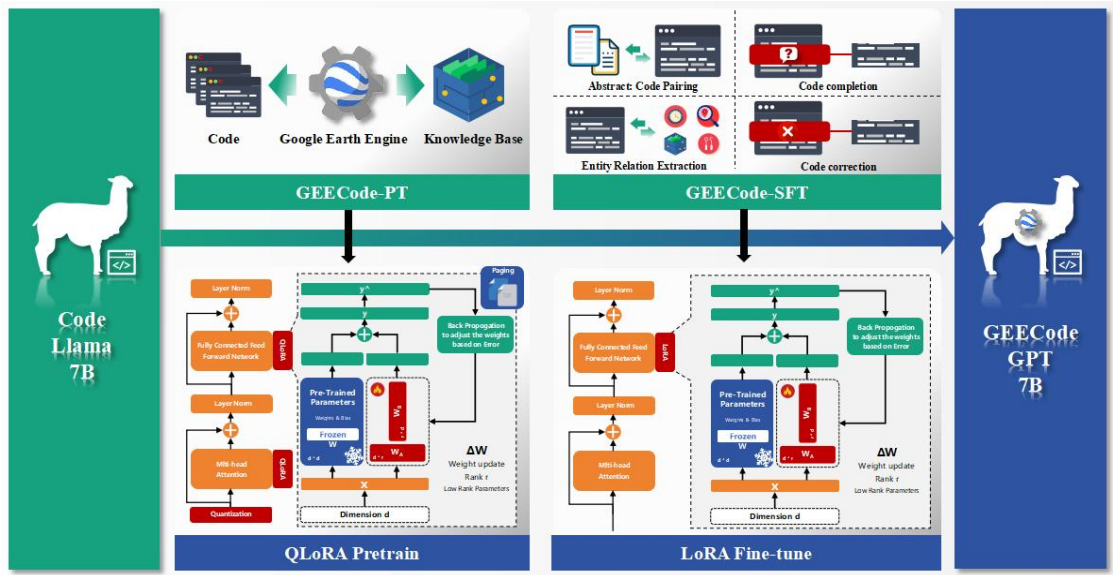

Figure.12 GEECode-GPT-7B Fine-tuning Diagram

In the pre-training phase, the QLoRA (Quantized Low-Rank Adaptation) method was used[89], which optimizes the decoder by combining quantization and low-rank adaptation, reducing memory consumption and improving computational efficiency. The hyperparameters were set as follows: global batch size of 64, random seed of 42, int4 quantization precision, a total of 1 training epoch, initial learning rate of 0.0002, using cosine decay with linear warmup for 5% of steps. Weight decay was set to 0.1, gradient accumulation was set to 4, and the maximum input length was 4096 tokens. The QLoRA rank value was 64, and the quantization type was NF4, with target modules including q_proj, v_proj, k_proj, o_proj, and MLP, and dropout probability set to 0.05. In the subsequent supervised fine-tuning stage, the LoRA(Low-Rank Adaptation) strategy was adopted[90], with an initial learning rate of 0.0001 and a batch size of 32, while keeping the gradient accumulation and input length settings the same. The LoRA rank value and target modules were consistent with pre-training, and fine-tuning was also performed using two NVIDIA A100 GPUs. The improvement data compared to some baseline models is shown in Table 12.

Table.12 Post-Fine-tuning Metrics Comparison

| Tasks | Indicators | GPT-4 | | LLaMA3-8B | | Code Llama-13B | | Code Llama-7B | | CodeGemma-7B | | CodeGeeX2-6B | | GEECode-GPT-7B |
|---|---|---|---|---|---|---|---|---|---|---|---|---|---|---|
| **Code Summarization** | Completeness | 0.924 | -0.164 | 0.812 | -0.052 | 0.52 | +0.240 | 0.556 | +0.204 | 0.468 | +0.292 | 0.384 | +0.376 | **0.760** |
| | Accuracy | 0.884 | -0.090 | 0.756 | +0.038 | 0.656 | +0.138 | 0.644 | +0.150 | 0.7 | +0.094 | 0.584 | +0.210 | **0.794** |
| | Text readability | 0.912 | -0.077 | 0.812 | +0.023 | 0.716 | +0.119 | 0.72 | +0.115 | 0.744 | +0.091 | 0.656 | +0.179 | **0.835** |
| | Overall | 0.907 | -0.110 | 0.793 | +0.004 | 0.631 | +0.166 | 0.640 | +0.157 | 0.637 | +0.160 | 0.541 | +0.256 | **0.797** |
| **Code Generation** | Executability | 0.488 | -0.078 | 0.244 | +0.166 | 0.36 | +0.050 | 0.32 | +0.090 | 0.22 | +0.190 | 0.308 | +0.102 | **0.410** |
| | Entity accuracy | 0.556 | -0.086 | 0.312 | +0.158 | 0.432 | +0.038 | 0.248 | +0.222 | 0.272 | +0.198 | 0.38 | +0.090 | **0.470** |
| | Code readability | 0.956 | -0.095 | 0.768 | +0.093 | 0.74 | +0.121 | 0.752 | +0.109 | 0.704 | +0.157 | 0.672 | +0.189 | **0.861** |
| | Overall | 0.667 | -0.086 | 0.441 | +0.139 | 0.511 | +0.070 | 0.440 | +0.140 | 0.399 | +0.182 | 0.453 | +0.127 | **0.580** |
| **Code Completion** | Executability | 0.744 | -0.082 | 0.596 | +0.066 | 0.564 | +0.098 | 0.548 | +0.114 | 0.520 | +0.142 | 0.504 | +0.158 | **0.662** |
| | Entity accuracy | 0.540 | -0.093 | 0.364 | +0.083 | 0.328 | +0.119 | 0.316 | +0.131 | 0.288 | +0.159 | 0.276 | +0.171 | **0.447** |
| | Code readability | 0.428 | -0.078 | 0.288 | +0.062 | 0.248 | +0.102 | 0.244 | +0.106 | 0.212 | +0.138 | 0.204 | +0.146 | **0.350** |
| | Overall | 0.571 | -0.084 | 0.416 | +0.070 | 0.380 | +0.106 | 0.369 | +0.117 | 0.340 | +0.146 | 0.328 | +0.158 | **0.486** |
| **Code Correction** | Executability | 0.600 | -0.064 | 0.444 | +0.092 | 0.564 | -0.028 | 0.404 | +0.132 | 0.392 | +0.144 | 0.380 | +0.156 | **0.536** |
| | Entity accuracy | 0.452 | -0.080 | 0.260 | +0.112 | 0.328 | +0.044 | 0.264 | +0.108 | 0.228 | +0.144 | 0.188 | +0.184 | **0.372** |
| | Code readability | 0.388 | -0.078 | 0.208 | +0.102 | 0.248 | +0.062 | 0.192 | +0.118 | 0.184 | +0.126 | 0.140 | +0.170 | **0.310** |
| | Overall | 0.480 | -0.074 | 0.304 | +0.102 | 0.380 | +0.026 | 0.287 | +0.119 | 0.268 | +0.138 | 0.236 | +0.170 | **0.406** |

After fine-tuning, the LLM exhibited significant improvement in subjective tasks such as summarization, code generation, code completion, and code correction in the Google Earth Engine platform's JavaScript environment, with enhancements ranging from 2.6% to 37.6% compared to general-purpose LLMs and general-purpose code generation models, achieving results close to the commercial LLM GPT-4. This demonstrates the feasibility of the fine-tuning approach.

## 6 Conclusion

This study proposed a novel evaluation task for LLMs and developed the GeoCode-Eval (GCE) framework. The framework comprehensively evaluates the geospatial code generation capability boundaries of LLMs across three dimensions—"Cognition and Memory", "Comprehension and Interpretation", and "Innovation and Creation"—and eight capability levels. The evaluated models included commercial LLMs, general-purpose LLMs, and code generation models, comparing different parameter scales of contemporary models. The evaluation framework has broad applicability. Based on the GCE framework, we collected extensive geospatial code, Wikipedia pages and introduction pages, built-in functions, syntax, and built-in datasets from open-source platforms, combined with expert experience and the Self-Instruct method to construct the GeoCode-Bench evaluation dataset. The evaluation set includes three types of objective questions (multiple-choice, fill-in-the-blank and true/false) and four types of subjective questions (code summarization, code generation, code completion and code correction). The evaluation adopted a combination of expert assessment and prompt engineering, incorporating multi-round voting and few-shot prompting to assess the impact of different prompt strategies on geospatial code generation performance. Finally, instruction fine-tuning was used to enhance the performance of LLMs, particularly with the GEECode-GPT-7B model derived from fine-tuning Code Llama-7B, which showed significant improvement.

Based on the exploration of LLMs' capability boundaries in geospatial code generation tasks, future research can further expand to explore other capability boundaries of LLMs in the geospatial domain. Additionally, examining ways to enhance LLM performance in geospatial code generation from both usage strategy optimization and model structure improvement perspectives holds significant research value. For geospatial code generation tasks, future work could focus on collecting larger and more representative geospatial domain corpora to build more reliable and highly specialized geospatial code generation models, thereby advancing research in this field.